\PassOptionsToPackage{sort&compress}{natbib}  
\documentclass[12pt]{elsarticle}
\usepackage{booktabs}
\usepackage{caption}

\usepackage{amsmath}
\usepackage{bm}
\usepackage[usenames, dvipsnames]{xcolor}
\usepackage[utf8x]{inputenc}
\usepackage[T1]{fontenc}
\usepackage{tikz}
\usepackage{pgfplots, pgfplotstable}
\usetikzlibrary{calc,arrows,fadings,decorations.pathreplacing,decorations.markings,patterns,shapes.geometric}
\usepackage{fp}
\usepackage{yhmath}
\usepackage{verbatim}
\usepackage{svg}
\usepackage{amssymb}
\usepackage{caption}
\usepackage{subcaption}
\usetikzlibrary{calc,fit,intersections,shapes}
\usetikzlibrary{calc}
\usepgfplotslibrary{polar}
\pgfplotsset{compat=1.6}
\pgfplotsset{every axis plot}
\usepackage{amsmath}
\usepackage{setspace}
\usepackage{lineno}

\usepackage{algorithm}
\usepackage{algpseudocode}

\usepackage{graphicx}
\usepackage{float}          
\graphicspath{{figuras/}}
\usepackage{siunitx}
\DeclareSIUnit{\pixel}{px}
\DeclareSIUnit{\gravity}{g}
\DeclareSIUnit{\fps}{fps}

\usepackage[hidelinks]{hyperref}
\usepackage{cleveref}

\newtheorem{proposition}{Proposition}
\newdefinition{remark}{Remark}

\newcommand{\ud}{\dot{\bm{u}}}
\newcommand{\udd}{\ddot{\bm{u}}}
\newcommand{\udk}[1]{\dot{\bm{u}}_{#1}}
\newcommand{\uk}[1]{\bm{u}_{#1}}
\newcommand{\ev}{\hat{\bm{e}}_{v}}
\newcommand{\ed}{\hat{\bm{e}}_{d}}
\newcommand{\ag}{\bm{a}_{g}}
\newcommand{\evk}[1]{\hat{\bm{e}}_{v,#1}}
\newcommand{\agk}[1]{\bm{a}_{g,#1}}

\begin{document}

\title{Rigid Bodies as Accidental Seismoscopes: Exact Time Stepper Inversion and an Estimate of Ground Acceleration in the 2026 La Guaira Earthquake}

\author[label1]{R. O. Ruiz \footnote{Corresponding author}}
\author[label2]{R. Boroschek}
\author[label3]{E. N\'u\~nez}

\address[label1]{Department of Mechanical Engineering, University of
Michigan-Dearborn, Dearborn, USA}
\address[label2]{Department of Civil Engineering, University of Chile,
Santiago, Chile}
\address[label3]{Department of Civil Engineering, Universidad Cat\'olica
de la Sant\'isima Concepci\'on, Concepci\'on, Chile}

\begin{singlespace}
\begin{abstract}

An unknown ground excitation is recovered from the measured response of a rigid body sliding in two dimensions on an inclined surface under Coulomb friction, a system with no classical solution, whose sampled trajectories come from a time stepper. The Moreau stepper is shown to admit an exact inverse, explicit, non-iterative and vectorizable, giving the ground acceleration over each step in closed form, without regularization or an assumed source model. The inverse evaluates the friction direction at the end of the step, not its beginning. Inverting the continuous equation instead is not a secondary discretization choice: the two inverses differ by an exact identity proportional to the rotation of the sliding direction within the step, bounded by twice the frictional deceleration, which is of order one $g$ for typical interfaces. The identity is inactive in one dimension, which explains why the sliding block literature has not confronted it. Verification against a manufactured reference solution recovers it to machine precision. On the field record, the discrepancy reaches \SI{0.10}{\gravity} in root mean square, which exceeds the whole propagated uncertainty of the reported peak, and it suppresses \SI{17}{\percent} of the recovered travel, behaving as a bias, not noise, so root-mean-square agreement in acceleration understates it. The method is applied to surveillance video of a delineator post that slid in the 24 June 2026 La Guaira earthquake, yielding a peak ground acceleration of \num{0.912}\,$\pm$\,\SI{0.047}{\gravity}, and a lower bound of \SI{0.47}{\gravity} independent of the imagery. Displaced rigid bodies can therefore provide quantitative ground-motion information where instruments are absent.

\end{abstract}
\end{singlespace}



\maketitle

\section{Introduction}
\label{sec:intro}

Estimating an unknown input from a measured output is a recurring task in mechanical signal processing, and it has been studied extensively in the context of force reconstruction, virtual sensing, and joint input-state estimation. For a linear time-invariant system, the problem takes a familiar form, i.e., the forward operator is a convolution, its inverse is a deconvolution, and the practical difficulty lies in the conditioning of that deconvolution and in selecting an appropriate regularization strategy. This formulation assumes the data were generated by a differential equation, so inverting the equation and inverting the process that produced the samples are the same operation.

This assumption does not hold for systems in unilateral contact or subject to a Coulomb friction law. Such systems do not possess classical solutions; the velocity is of bounded variation, the contact force is a measure, and the equation of motion is satisfied only in an integrated sense or in the sense of a differential inclusion \citep{moreau1988,stewart2000}. The practical consequence, and the starting point of the present work, is that the object that actually produces any given sampled trajectory is a time-stepping scheme. Moreau's scheme and its relatives \citep{jean1999,anitescu1997,acary2008,studer2009} are not approximations converging to a classical solution of a differential equation, since no such solution exists; they define the solution. It follows that when such a system is inverted, the object that should be inverted is the scheme rather than the equation that motivated it.

That the discrete and the continuous descriptions need not be interchangeable in the presence of contact is not, in itself, a new observation. This is well known in the differentiable simulation literature, where gradients are propagated through frictional contact solvers for system identification and design. Geilinger et al. \citep{geilinger2020} address the non-differentiability of normal and tangential contact forces by a principled mollification combined with fully implicit integration, and Li et al. \citep{li2022diffcloth} analyze the sources of non-differentiability under the Signorini-Coulomb law and derive the Jacobian of the discrete update rather than of the continuous model. The recognition in that community is qualitative and general: differentiate then discretize and discretize then differentiate do not commute once contact is present, and the discrete route is the one consistent with the computed trajectory. What appears to be missing is a setting in which the two routes can be compared exactly rather than numerically, in which the discrepancy between them can be written in closed form, and in which its magnitude can be shown to be of engineering rather than of academic size. The present paper provides such a setting.

The system chosen for this purpose is a rigid body sliding in two dimensions on an inclined plane under Coulomb friction, with unknown ground acceleration and measured relative displacement. Its forward counterpart is the sliding block of \citet{newmark1965}, which predicts permanent displacement from a known accelerogram and which supports a substantial body of work in geotechnical earthquake engineering \citep{jibson2007,rathje2000}. Refinements of the forward model have considered coupled and nonlinear formulations \citep{rathje2000}, and the effect of treating the excitation as genuinely multi-directional rather than as a single horizontal component has been examined as well \citep{rathje2001directionality}. These studies focus on the accuracy of the predicted displacement. The inverse version of the problem, in which the displacement is known and the accelerogram is the unknown, has received considerably less attention, and that attention has been predominantly one-dimensional.

A related line of work has developed in seismology and in forensic engineering, where the objective is to constrain the ground motion of a past event from the observed state of objects that were present during it. Researchers have treated precariously balanced rocks as upper-limit strong-motion seismoscopes, established their overturning fragilities numerically, and validated them on shaking tables \citep{anooshehpoor2004,purvance2008,zhangmakris2001}. They have also characterized the rocking response of free-standing blocks in considerable detail. A general framework for inferring ground motion levels from earthquake damage has also been proposed \citep{forensicseismology2014}, and non-instrumental evidence has been used to constrain exceptional near-source motions when instrumental coverage was inadequate \citep{leteil2020}. Two features characterize this literature. The observable is almost always rocking or overturning rather than sliding, and the product is almost always a bound or a fragility rather than a recovered time history.

More recently, the increasing availability of video of the built environment has motivated the direct use of imagery for this purpose. Vision-based displacement measurement is now well established in structural health monitoring, and several authors have extended it to identify the excitation itself \citep{boroschek2025nonlinear}. Methods have been proposed to extract structural seismic response from blurred surveillance video by line matching \citep{blurredvideo2025}, to combine recorded video with accelerometric data in order to determine nonlinear response and residual drift \citep{boroschek2025nonlinear}, and to exploit surveillance and personal cameras as a distributed observational resource for the earthquake wave field \citep{surveillancewavefield2021}. Closest to the present work, a fully vision-based framework has been proposed for the identification of ground motion time histories from building vibration video, with validation on a five-story shear building on a shaking table \citep{visionGM2025}. That framework inverts a linear structural model: the observed system is compliant, its response is governed by a differential equation, and classical structural dynamics supplies the inverse. The present work addresses the complementary situation, in which the observed system is a rigid body in frictional contact, no differential equation governs it, and the inverse must therefore be constructed from the time stepper.

Motivated by these observations, this work aims to formulate and verify an exact inverse for the two-dimensional Coulomb slider, along with a quantitative account of what is lost when the continuous equation is inverted in its place. Five specific results are reported. First, it is shown that the Moreau time stepper for this system admits an exact inverse that is explicit, non-iterative and fully vectorizable, and that requires no regularization, no assumed spectral content and no fitted source model. Second, it is shown that the friction direction in this inverse must be evaluated at the end of the step rather than at its beginning, that the implicitness of this direction is precisely the mechanism by which the scheme arrests a sliding body. Additionally, the discrepancy introduced by evaluating it at the beginning of the step instead is given by an exact identity whose bound is twice the frictional deceleration and therefore of order one $g$ for typical interfaces. Third, we show that this identity is inactive in one dimension, where the sliding direction takes only two values and changes only at velocity reversals, which could explain why an effect of this magnitude has remained unremarked within a model in use for six decades. Fourth, the exact inverse is verified against a manufactured reference solution, constructed by prescribing the relative motion analytically and defining the excitation as its exact inverse, so that the reference is exact by construction rather than accurate to the order of an integration scheme. We further show that the discrepancy behaves as a bias, not noise, with consequences for how inversions of this class should be validated. Fifth, we apply the framework to a field recording (case study) and report the complete chain from video to ground acceleration.

The case study deserves a word of justification, since it carries more weight here than an illustrative example normally would. On 24 June 2026, two earthquakes of magnitude 7.2 and 7.5 occurred approximately \SI{39}{\second} apart on the Venezuelan Caribbean coast, producing the most destructive seismic event in that country in more than a century, with the heaviest damage concentrated in the state of La Guaira \citep{usgs2026laguaira}. Near-source instrumental coverage of the affected region is sparse; consequently, reconstructing the ground motion that produced the observed damage depends heavily on estimates derived from non-instrumental evidence. In such a setting, any independent, quantitative constraint on peak ground acceleration directly informs the seismological and engineering characterization of the event, and observations that would ordinarily be regarded as anecdotal gain evidential weight. The recording analyzed here is of this kind: a fixed surveillance camera that happened to view a delineator post which slid during the strong motion. We treat it accordingly and document the full measurement chain so the resulting estimate can be assessed, reproduced, and, if appropriate, superseded.

The rest of the paper is organized as follows. Section~\ref{sec:formulacion} establishes the forward time-stepping scheme, derives the exact inverse, and obtains the identity relating it to the inverse of the continuous equation. Section~\ref{sec:gemelo} constructs the manufactured reference solution, reports the verification and the parametric sweeps, and examines the distinction between bias and noise. Section~\ref{sec:aplicacion} presents the case study, following the measurement chain from the raw recording to the estimated peak ground acceleration and its uncertainty. Section~\ref{sec:conclusiones} discusses the scope of the results, their limitations, and the experimental work that would be required to remove them.

\section{Dynamics of Sliding Bodies}
\label{sec:formulacion}

\subsection{Equation of Motion}
\label{sec:sistema}

Throughout this paper, boldface symbols denote planar vectors and italic symbols denote scalars. A hat denotes a unit-magnitude vector, vertical bars denote the Euclidean magnitude of a vector, an overdot denotes differentiation with respect to time, and a subscript $i$ denotes evaluation at the discrete instant $t_{i}$. The relative kinematics are expressed exclusively through the displacement $\bm{u}$ and its time derivatives $\ud$ and $\udd$, and no separate symbol is introduced for the relative velocity.

Consider a rigid body of mass $m$ resting on a plane inclined by a small angle $\theta$ from the horizontal, in contact through a dry frictional interface of kinetic coefficient $\mu_{k}$ and static coefficient $\mu_{s}$, with $\mu_{s}\ge\mu_{k}$. The plane translates with an unknown acceleration $\ag(t)$ and does not rotate, and $g$ denotes the acceleration of gravity. Let $\bm{u}(t)$ denote the displacement of the body relative to the plane, resolved in a frame attached to the plane, so that $\ud(t)$ is the relative velocity and $\udd(t)$ the relative acceleration.

Two unit vectors appear in the formulation and must be kept distinct, since they play different roles and only one of them is constant. The first, $\ed$, is the downslope direction of the plane; it is fixed, and it multiplies the component of gravity acting along the surface, $g\sin\theta$. The second, $\ev=\ud/\lvert\ud\rvert$, is the instantaneous direction of sliding (it rotates as the body moves), and it multiplies the frictional deceleration $\mu_{k}g\cos\theta$. The subscript $v$ is retained as a mnemonic for the velocity direction. Conflating these two vectors is harmless in one dimension but incorrect in two, and the results reported in this paper arise entirely from the rotation of $\ev$.

In the sliding regime, the balance of momentum, written in the non-inertial frame of the plane and divided through by the mass $m$, corresponds to
\begin{equation}
  \udd
  \;=\;-\,\ag
  \;+\;g\sin\theta\,\ed
  \;-\;\mu_{k}\,g\cos\theta\,\ev ,
  \label{eq:eom}
\end{equation}
where the three terms on the right-hand side are, respectively, the inertial forcing from the plane's motion, the downslope component of gravity, and the frictional deceleration opposing sliding. The body remains stationary as long as the tangential force required to hold it in place lies within the friction cone, that is, as long as
\begin{equation}
  \bigl\lvert -\ag+g\sin\theta\,\ed \bigr\rvert
  \;\le\;\mu_{s}\,g\cos\theta ,
  \label{eq:stick}
\end{equation}
where the left-hand side is the magnitude of the net non-frictional tangential demand and the right-hand side is the largest tangential resistance the interface can supply. We take the normal load as $mg\cos\theta$ throughout, assuming contact is maintained and no overturning moment develops.

During sliding, the relative motion provides two scalar equations, while the unknowns are the two components of $\ag$, $\mu_{k}$, $\theta$, and the azimuth of $\ed$. The problem is therefore underdetermined by three, and no amount of kinematic processing will close it: those three numbers must come from elsewhere. Given them, however, the inversion is exact and pointwise. There is no regularization parameter to choose and no trade-off curve to draw, which is unusual for an inverse problem and follows directly from the friction law being algebraic rather than dynamic.

One statement survives independently of all three unknowns. The body moved, so at some instant \cref{eq:stick} must have been violated, and since $\mu_{s}\ge\mu_{k}$ always,
\begin{equation}
  \max_{t}\lvert\ag(t)\rvert
  \;\ge\;\mu_{k}\,g\cos\theta .
  \label{eq:cota}
\end{equation}
This bound uses the kinematics only through the binary fact that motion occurred. Everything that follows should be read as a refinement of \cref{eq:cota} and not as a substitute for it.

\subsection{Time Stepper for the Forward Problem}
\label{sec:adelante}

Equation~\eqref{eq:eom} has no classical solution across the transitions between sticking and sliding: $\ev$ is undefined when $\ud=\bm{0}$, and the friction force there is set-valued. The standard formulation replaces \cref{eq:eom} by a differential inclusion and integrates it with a time-stepping scheme of Moreau type \citep{moreau1988,jean1999,acary2008}. Over a step of length $\Delta t$ from $t_{i}$ to $t_{i+1}$, with $\udk{i}$ known, the scheme takes a trial velocity that ignores friction,
\begin{equation}
  \ud^{\ast}
  \;=\;\udk{i}
  \;+\;\Delta t\,\bigl(-\agk{i}+g\sin\theta\,\ed\bigr),
  \label{eq:trial}
\end{equation}
and then applies the friction impulse through the proximal operator associated with the friction cone, which for an isotropic Coulomb law and a scalar normal load reduces to a radial shrinkage,
\begin{equation}
  \udk{i+1}
  \;=\;\ud^{\ast}\,
  \max\!\left(0,\;
  1-\frac{\Delta t\,\mu_{k}\,g\cos\theta}{\lvert\ud^{\ast}\rvert}\right),
  \qquad
  \uk{i+1}\;=\;\uk{i}+\Delta t\,\udk{i+1}.
  \label{eq:moreau}
\end{equation}
The scheme is unconditionally stable, it sticks exactly rather than asymptotically, and it requires no event detection: if the trial velocity is smaller than the friction impulse, the shrinkage factor is clipped at zero, and the body stops within the step. These are the properties that make Moreau stepping the standard tool for this class of system, and they are properties of \cref{eq:moreau}, not of \cref{eq:eom}. The following elementary observation is the hinge of the paper.

\begin{proposition}[Direction preservation]
\label{prop:direccion}
The factor multiplying $\ud^{\ast}$ in \cref{eq:moreau} is a non-negative scalar. Hence $\udk{i+1}$ is parallel to $\ud^{\ast}$ whenever it is non-zero, and the friction impulse over the step is directed along $\evk{i+1}$, the sliding direction at the \emph{end} of the step.
\end{proposition}

\noindent The proof is the statement. This implies that the discrete friction direction is implicit, and this is not an aesthetic preference among discretizations. Then, explicit friction and exact sticking are incompatible. Suppose we took the direction explicitly along $\evk{i}$. A body arriving at the step with a small but non-zero velocity would then receive the full impulse $\Delta t\,\mu_{k}g\cos\theta$ along the direction it came from, overshoot zero, and leave the step moving backward. Repeating, it oscillates about rest with an amplitude set by $\Delta t$. This is the classical chattering of regularized and explicit Coulomb models, and it is why such models require either an artificial velocity threshold or stiff regularization, both of which introduce a parameter with no physical counterpart. The implicitness of the direction in \cref{eq:moreau} \emph{is} the mechanism by which the scheme sticks.

\subsection{The Exact Inverse of the Time Stepper}
\label{sec:inverso}

We now invert \cref{eq:moreau} rather than \cref{eq:eom}. Suppose the relative velocity is known at the grid points, $\udk{0},\dots, \udk{N}$, and that step $i$ is a sliding step, meaning $\udk{i+1}\neq\bm{0}$. By \cref{prop:direccion} the trial velocity is recovered from the end-of-step state without ambiguity,
\begin{equation}
  \ud^{\ast}
  \;=\;\udk{i+1}
  \;+\;\Delta t\,\mu_{k}\,g\cos\theta\;\evk{i+1},
  \qquad
  \evk{i+1}=\frac{\udk{i+1}}{\lvert\udk{i+1}\rvert},
  \label{eq:trialinv}
\end{equation}
because the shrinkage in \cref{eq:moreau} scales the magnitude and leaves the direction alone, so $\lvert\ud^{\ast}\rvert= \lvert\udk{i+1}\rvert+\Delta t\,\mu_{k}g\cos\theta$. Substituting \cref{eq:trialinv} into \cref{eq:trial} and solving for the acceleration gives the result.

\begin{proposition}[Exact discrete inverse]
\label{prop:inverso}
For every sliding step, the ground acceleration is given explicitly by \cref{eq:inverso}. Re-integrating \cref{eq:moreau} with the sequence $\{\agk{i}\}$ so obtained reproduces $\{\udk{i}\}$ and $\{\uk{i}\}$ exactly, in the sense of floating-point arithmetic.
\end{proposition}
\begin{equation}
  \;
  \agk{i}
  \;=\;g\sin\theta\,\ed
  \;-\;\frac{1}{\Delta t}
  \Bigl[
  \bigl(\udk{i+1}+\Delta t\,\mu_{k}g\cos\theta\,\evk{i+1}\bigr)
  -\udk{i}
  \Bigr].
  \label{eq:inverso}
\end{equation}

Three features of \cref{eq:inverso} deserve emphasis, because they are the practical payoff of inverting the scheme rather than the equation. First, it is explicit. No iteration, root finding, or complementarity solve is required, in contrast to the forward step, which involves a projection. Second, it is local. Step $i$ uses only $\udk{i}$ and $\udk{i+1}$, so errors do not propagate along the record, and it is therefore vectorizable, unlike the forward recursion. And, third, it carries no regularization. The inversion is a rearrangement of the scheme, not the minimization of a misfit.

The inverse is silent on the stuck steps, where $\udk{i+1}=\bm{0}$ and $\evk{i+1}$ is undefined. There \cref{eq:stick} bounds $\ag$ but does not determine it, and the correct report is an interval rather than a value. This is a genuine loss of information and not an artifact of the method: a body at rest cannot distinguish among sub-threshold excitations. We note in \cref{sec:aplicacion} that, for the record analyzed there, the loss does not affect the quantity of interest.

\subsection{The Index Identity}
\label{sec:identidad}

Suppose now that one proceeds in the natural but incorrect way: discretize \cref{eq:eom} directly by a forward difference and evaluate the friction direction at the beginning of the step, obtaining
\begin{equation}
  \tilde{\ag}_{,i}
  \;=\;g\sin\theta\,\ed
  \;-\;\frac{1}{\Delta t}
  \Bigl[
  \bigl(\udk{i+1}+\Delta t\,\mu_{k}g\cos\theta\,\evk{i}\bigr)
  -\udk{i}
  \Bigr].
  \label{eq:ingenuo}
\end{equation}
Comparing \cref{eq:inverso,eq:ingenuo}, everything cancels except the friction term, and the discrepancy is not an order-of-magnitude estimate but an identity:
\begin{equation}
  \Delta\agk{i}
  \;\equiv\;\agk{i}-\tilde{\ag}_{,i}
  \;=\;\mu_{k}\,g\cos\theta\,
  \bigl(\evk{i+1}-\evk{i}\bigr),
  \qquad
  \lvert\Delta\agk{i}\rvert\;\le\;2\,\mu_{k}\,g\cos\theta .
  \label{eq:identidad}
\end{equation}
The bound is attained on a reversal and is independent of $\Delta t$, the amplitude of the motion, and the excitation. For the interface measured in \cref{sec:aplicacion}, $\mu_{k}=\num{0.470}$ and $\theta=\ang{2.38}$, it equals \SI{0.940}{\gravity}, which is of the same order as the peak acceleration being sought.

In one-dimensional problems, the motion is confined to a line, $\ev\in\{-1,0,+1\}$, and the right-hand side of \cref{eq:identidad} is zero at every step that is not a reversal. Substituting $\evk{i}$ for $\evk{i+1}$ is then harmless almost everywhere, and the error it produces is confined to isolated instants and could be misread as a transition artifact. In two dimensions $\ev$ rotates continuously, so \cref{eq:identidad} is active at essentially every step, and the error is not confined to transitions at all. We quantify that statement numerically in \cref{sec:costo}, where the strong-motion window of the field record contains a single stick-slip transition and removing it leaves \SI{97}{\percent} of the discrepancy intact.

Under the particular case in which the sliding direction rotates at a constant rate $\Omega_{v}$, so that the angle between $\evk{i}$ and $\evk{i+1}$ is $\Omega_{v}\Delta t$, the chord between two unit vectors gives
\begin{equation}
  \lvert\Delta\agk{i}\rvert
  \;=\;2\,\mu_{k}\,g\cos\theta\,
  \sin\!\left(\frac{\Omega_{v}\,\Delta t}{2}\right).
  \label{eq:cerrada}
\end{equation}
\Cref{eq:cerrada} is verified numerically to \SI{3.3e-16}{\gravity} in \cref{sec:barridos}. It makes three things immediate. The error grows monotonically with the rotation rate and saturates at the bound $2\mu_{k}g\cos\theta$ when the direction reverses within a single step. It is first order in $\Delta t$ for slow rotation, since \cref{eq:cerrada} tends to $\mu_{k}g\cos\theta\,\Omega_{v}\Delta t$. And it is exactly proportional to $\mu_{k}$, so it cannot be made small by appealing to a low-friction interface unless the friction itself is negligible, in which case there is no sliding problem to solve.

It is tempting to dismiss \cref{eq:identidad} as an $O(\Delta t)$ effect that refinement will remove. That argument fails whenever $\Delta t$ is set by the instrument rather than by the analyst. In the application of \cref{sec:aplicacion} the sampling interval is fixed at \SI{40}{\milli\second} by a \SI{25}{\fps} camera, the velocity increment $\Delta t\,\mu_{k}g\cos\theta=\SI{184}{\milli\meter\per\second}$ is \SI{15}{\percent} of the peak sliding speed, and \cref{sec:barridos} shows that suppressing the discrepancy below \SI{0.05}{\gravity} would require \SI{41}{\fps}. That rate is modest by the standards of a laboratory and unavailable from the hardware that produced this record, which is the situation of any fixed-rate acquisition the analyst did not specify, and it tightens in proportion to the rotation rate of the sliding direction, so a record that turns faster than this one demands more. Where refinement \emph{is} available, \cref{eq:inverso} removes the error at fixed $\Delta t$ and at no cost, so the argument for using it does not rest on the sampling rate being coarse.

\section{Verification Against a Manufactured Reference Solution}
\label{sec:gemelo}

A claim of exactness cannot be supported by an application, since in an application the quantity that the inverse is supposed to return is not known independently. It is therefore necessary to construct a setting in which it is known, and the construction adopted here is the method of manufactured solutions, which is standard practice in the verification of computational models \citep{roache1998}. In that method one does not attempt to solve the governing problem and then assess the answer. One instead prescribes the response, substitutes it into the governing relations, and defines the forcing to be whatever those relations then require. The resulting pair of response and forcing is exact by construction, and any departure observed when the forcing is recovered from the response is attributable to the recovery procedure alone.

It should be pointed out that this is not a digital twin in the sense that term has acquired in the condition monitoring literature, where it denotes a model kept in bidirectional correspondence with an operating asset. No such correspondence is involved here. What is constructed is a reference solution whose excitation is known analytically, and it is used for verification in the strict sense, that is, to establish that the inverse operator returns what it was derived to return.

\subsection{Prescribed Velocity}
\label{sec:construccion}

The natural way to construct such a reference is to prescribe a ground acceleration, integrate \cref{eq:moreau} forward, and then attempt to recover the acceleration from the resulting motion. This route was considered and discarded for three reasons. The first is that the prescribed acceleration is then known only to the accuracy with which it has been represented on the time grid, so that the reference is itself approximate and the verification is limited by it. The second is that the rotation rate of $\ev$, which \cref{eq:cerrada} identifies as the quantity governing the discrepancy between the two inverses, is an outcome of the integration rather than an input to it, and therefore cannot be prescribed or swept. The third is that sticking occurs wherever the prescribed excitation happens to fall below the threshold of \cref{eq:stick}, which is undesirable in a study whose purpose is to establish that the effect under examination is not a transition effect.

The construction is therefore inverted. The relative velocity is
prescribed analytically as
\begin{equation}
  \ud(t)\;=\;V(t)\,
  \bigl[\cos(\Omega_{v}t+\phi),\;\sin(\Omega_{v}t+\phi)\bigr],
  \qquad
  V(t)\;=\;\bar{V}\bigl[1+\beta\sin(2\pi f t)\bigr],
  \label{eq:prescrita}
\end{equation}
where $V(t)$ denotes the sliding speed, $\Omega_{v}$ the angular rate at which the sliding direction $\ev$ rotates, $\phi$ its initial orientation, $\bar{V}$ the mean sliding speed, $\beta$ the relative amplitude of the speed modulation, and $f$ its frequency. The modulation amplitude is restricted to $\beta<1$ so that $V(t)>0$ throughout and the body never comes to rest. This velocity is sampled on the time grid, and the true ground acceleration is then \emph{defined} to be the exact inverse \cref{eq:inverso} of the sampled velocity. Three properties follow from this definition.

The first is that the reference solution is exactly self-consistent. Integrating \cref{eq:moreau} forward with the acceleration so defined returns the prescribed velocity, and does so without any accuracy caveat, because the inequality
\begin{equation}
  \lvert\ud^{\ast}\rvert
  \;=\;\lvert\udk{i+1}\rvert+\Delta t\,\mu_{k}g\cos\theta
  \;>\;\Delta t\,\mu_{k}g\cos\theta
  \label{eq:consistencia}
\end{equation}
holds identically, where $\ud^{\ast}$ is the unconstrained velocity predictor of \cref{eq:moreau} and the two sides are compared in magnitude. The shrinkage factor in \cref{eq:moreau} is consequently never clipped at zero, and the proximal operator never removes more momentum than the frictional impulse of the step can account for. Numerically the round trip closes to \SI{2.3e-13}{\milli\meter\per\second} in velocity and \SI{5.7e-14}{\milli\meter} in displacement, which is the double-precision floor for quantities of this magnitude.

The second is that the rotation rate $\Omega_{v}$ has become a direct experimental variable, which is what \cref{eq:cerrada} requires if its prediction is to be tested rather than merely observed. The third is that sticking has been excluded by construction, so that any discrepancy observed in this reference solution cannot be attributed to stick-slip transitions.

The interface parameters used throughout this section are those measured in \cref{sec:aplicacion}, namely $\mu_{k}=\num{0.470}$, $\theta=\ang{2.38}$ and a downslope azimuth of \ang{161.8}. These values are not arbitrary and are not chosen for convenience. They are the values carried over from the case study, so that the magnitudes reported here are the magnitudes that the field application will encounter rather than magnitudes selected to make the effect appear large. All three are measurements: the friction coefficient on the bench in \cref{sec:friccion}, the inclination and the azimuth by inclinometer survey in \cref{sec:pendiente}. None of the conclusions of this section depends on the azimuth in any case, since it enters only through the downslope unit vector $\ed$ and not through the identity \cref{eq:identidad}. With these values the frictional deceleration is $\mu_{k}g\cos\theta=\SI{4606.9}{\milli\meter\per\second\squared} =\SI{0.4696}{\gravity}$ and the bound of \cref{eq:identidad} is \SI{0.9392}{\gravity}. Where a parameter is swept, it is swept about the measured value, and the measured value is marked on the corresponding figure.

\subsection{Verification on a Short Record}
\label{sec:demo}

\begin{figure}[H]
  \centering
  \includegraphics[width=\textwidth]{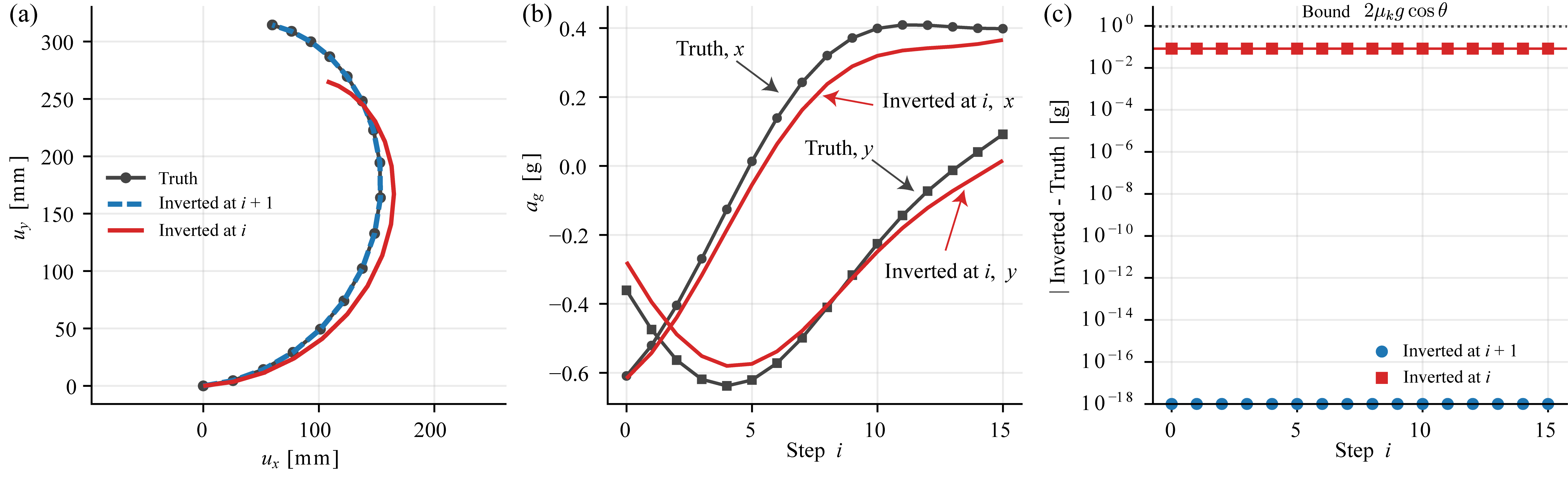}
  \caption{Verification on a \num{16}-step manufactured reference solution with no noise and no sticking. (a) Trajectory obtained by re-integrating   \cref{eq:moreau} with each recovered acceleration; the exact inverse is   indistinguishable from the reference, whereas the inverse taken at step   $i$ falls short. (b) Recovered ground acceleration components. (c) Error   per step on a logarithmic scale: the exact inverse lies at the floating-point floor, the inverse at $i$ sits at \SI{0.0827}{\gravity}, which is the closed form \cref{eq:cerrada} evaluated at this rotation rate, and the upper dotted line is the bound $2\mu_{k}g\cos\theta$.}
  \label{fig:demo}
\end{figure}

\Cref{fig:demo} reports a \num{16}-step reference solution at $\Delta t= \SI{40}{\milli\second}$ and $\Omega_{v}=\SI{4.41}{\radian\per\second}$, corresponding to a rotation of the sliding direction of \ang{10.10} per step. The time step is the frame interval of the recording analyzed in \cref{sec:aplicacion} and the rotation rate is the mean rotation measured over the strong-motion window of that record, so that the verification is performed at the operating point of the application rather than at a point chosen to make the effect large. The sliding speed ranges from \SIrange{438}{810}{\milli\meter\per\second} and no step sticks.

The exact inverse recovers the acceleration with a maximum error of \SI{0}{\milli\meter\per\second\squared}, that is, identically zero in double precision, and the identity \cref{eq:identidad} is satisfied to a residual of \SI{2.5e-12}{\milli\meter\per\second\squared}. The inverse evaluated at step $i$ is in error by \SI{0.0827}{\gravity} at every step, which matches \cref{eq:cerrada} to four significant figures. Note that the error is constant precisely because the rotation rate has been prescribed to be constant, and that no transition occurs anywhere in the record, so that the discrepancy cannot be a transition effect.

The consequence for the quantity that would ordinarily be reported is direct. Re-integrating the forward scheme with each recovered acceleration, the reference travel over the window is \SI{320.06}{\milli\meter}, the exact inverse returns \SI{320.06}{\milli\meter} or \SI{100.0}{\percent}, and the inverse at $i$ returns \SI{286.05}{\milli\meter} or \SI{89.4}{\percent}. A tenth of the displacement is lost through a change of index, at a rotation rate that was not chosen but measured.

\subsection{Parametric Sweeps}
\label{sec:barridos}

\begin{figure}[H]
  \centering
  \includegraphics[width=\textwidth]{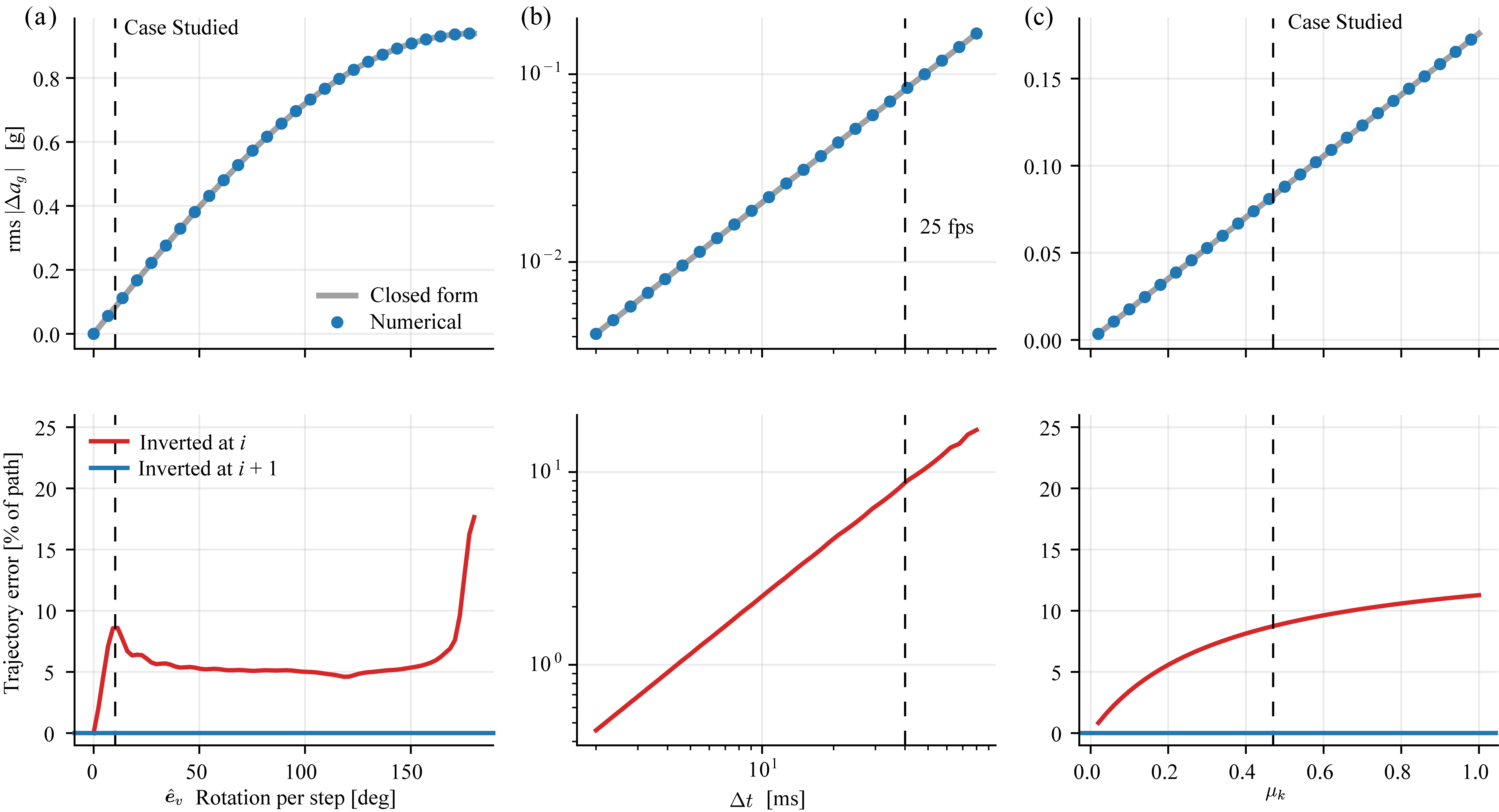}
  \caption{Parametric sweeps with no noise. Top row: root-mean-square error of the recovered ground acceleration for the inverse taken at step   $i$, with the closed form \cref{eq:cerrada} overlaid. Bottom row: root-mean-square trajectory departure as a percentage of the reference  arc length; the exact inverse is the flat line at zero. (a) Rotation rate of $\ev$. (b) Time step, with the physics held fixed. (c) Friction coefficient. Dashed vertical lines mark the values of the field record.}
  \label{fig:barridos}
\end{figure}

\Cref{fig:barridos} sweeps the three quantities that \cref{eq:cerrada} predicts should govern the discrepancy, one at a time, with the remaining quantities held at the values stated in \cref{sec:construccion}. The upper row reports the root-mean-square error in the recovered acceleration and the lower row reports the root-mean-square departure of the re-integrated trajectory from the reference, normalized by the arc length of the reference path. It should be pointed out that the trajectory metric is normalized by arc length rather than expressed as a ratio of endpoint displacements, because at high rotation rates the net displacement approaches zero and any endpoint ratio becomes ill-conditioned.

Regarding the rotation of the sliding direction. Sweeping $\Omega_{v}$ over the full admissible range, from no rotation to a reversal of the sliding direction within a single step, the numerical error agrees with \cref{eq:cerrada} to a maximum absolute discrepancy of \SI{3.331e-16}{\gravity}, which is the double-precision floor. The curve saturates at \SI{0.9392}{\gravity}, equal to the bound $2\mu_{k}g\cos\theta$ to every digit reported. At the rotation rate of the field record, \ang{10.1} per step, the error is \SI{0.0827}{\gravity} and the trajectory departs by \SI{8.7}{\percent} of the arc length. The exact inverse remains at zero across the whole sweep.

Regarding the time step. Holding the physics fixed and refining $\Delta t$, the observed convergence order is \num{0.9991}, which confirms that the discrepancy is first order in the time step and therefore a genuine consistency error rather than an implementation artifact. The practical reading of this panel is less comfortable than the convergence order alone would suggest. At $\Delta t=\SI{40}{\milli\second}$, the frame interval of the recording, the error is \SI{0.0827}{\gravity} and the trajectory departs by \SI{8.7}{\percent}. Reaching \SI{0.05}{\gravity} would require $\Delta t=\SI{24.2}{\milli\second}$, that is \SI{41}{\fps}. That rate is not delivered by surveillance hardware, which is the case of interest here, and the requirement tightens in proportion to the rotation rate, so a record whose sliding direction turns faster than this one would demand more. The deeper objection is that refinement is a remedy only where refinement is available, and for an incidental recording it never is, whereas \cref{eq:inverso} is a remedy at any time step.

Regarding the Friction coefficient. The error is linear in $\mu_{k}$ with a slope of \SI{0.1759}{\gravity} per unit $\mu_{k}$ and an intercept of $\num{-1.9e-19}$, in agreement with the proportionality stated in \cref{eq:cerrada}. At the measured value the error is \SI{0.0827}{\gravity}. There is accordingly no low-friction regime in which the question becomes immaterial, short of the regime in which the frictional resistance is so small that there is no sliding problem to speak of.

\subsection{A Reference Solution Built on Measured Kinematics}
\label{sec:realista}

The synthetic reference solution of \cref{sec:construccion} is clean by design, which is both its purpose and its limitation: its sliding direction rotates at a constant rate and the body never comes to rest, whereas a field record does neither. A second reference solution is therefore constructed whose kinematics are those actually measured in the field. The tracked relative displacement obtained in \cref{sec:aplicacion} is smoothed with a Savitzky-Golay filter \citep{savitzky1964}, differenced to give a velocity, and thresholded to identify the intervals in which the body is stationary. The resulting velocity sequence is then declared to be the reference, and the reference acceleration is defined as its exact inverse, exactly as in \cref{sec:construccion}. The construction thereby retains the stick-slip structure, the rotation history and the amplitude of the real event while preserving a known answer.

This reference solution comprises \num{361} frames, of which \num{72} involve sliding (\SI{20}{\percent}), with a total travel of \SI{422.2}{\milli\meter}, a peak reference acceleration of \SI{0.9332}{\gravity} and a mean rotation of $\ev$ of \ang{10.2} per step taken over the steps of sustained sliding, with a median of \ang{7.6} and a largest value of \ang{47.1}. These are not selected values but the values that the field record supplies, and they are reported here so that the numbers appearing in the remainder of this section can be read against the case study of \cref{sec:aplicacion} directly. That mean rotation is the same as the \ang{10.1} which \cref{sec:costo} measures over the strong-motion window alone, the two populations differing not in their mean but in their spread: the window contains no step turning by more than \ang{22.2}, whereas the record as a whole does. Forward consistency closes to \SI{2.8e-13}{\milli\meter}.

Taken over all sliding steps, the inverse at step $i$ is in error by \SI{0.2025}{\gravity} in root mean square. That aggregate should not be decomposed naively, because the two kinds of sliding step are not commensurable. At the \num{10} steps on which the body departs from rest the sliding direction at the beginning of the step does not exist, the body being stationary there, so the inverse taken at step $i$ has no defined value and the convention used here, which sets the direction to zero, makes its error exactly $\mu_{k}g\cos\theta=\SI{0.4696}{\gravity}$ by construction. Those steps are not evidence about the index. They are evidence that the inverse of the continuous equation is undefined at departure, which is a separate and stronger objection to it.

The informative population is therefore the \num{62} steps of sustained sliding, on which the direction exists at both ends and the two inverses can be compared at all. Over those steps the discrepancy is \SI{0.1098}{\gravity}, obtained with no transition anywhere in it, and it is more than twice the entire uncertainty attached to the recovered peak in \cref{eq:pga}. The same decomposition carried out on the field record itself is reported in \cref{sec:costo}, where the strong-motion window contains a single transition and its removal leaves \SI{97}{\percent} of the discrepancy in place. By either route the conclusion is the one stated in \cref{sec:identidad}: the discrepancy is a property of sustained sliding, during which the direction rotates, and not of the instants of departure and arrest, which is the attribution that one-dimensional intuition would naturally suggest. The travel recovered with the inverse at $i$ is \SI{360.9}{\milli\meter}, or \SI{85.5}{\percent}.

\subsection{Bias and Noise Effect}
\label{sec:ruido}

Field kinematics are measured, and measurement introduces noise. It might reasonably be expected that tracking noise, being the larger error source in practice, would render the distinction between the two inverses immaterial. It does not, and the reason is worth stating with some care, since it bears directly on how inversions of this class ought to be
validated.

\begin{figure}[H]
  \centering
  \includegraphics[width=\textwidth]{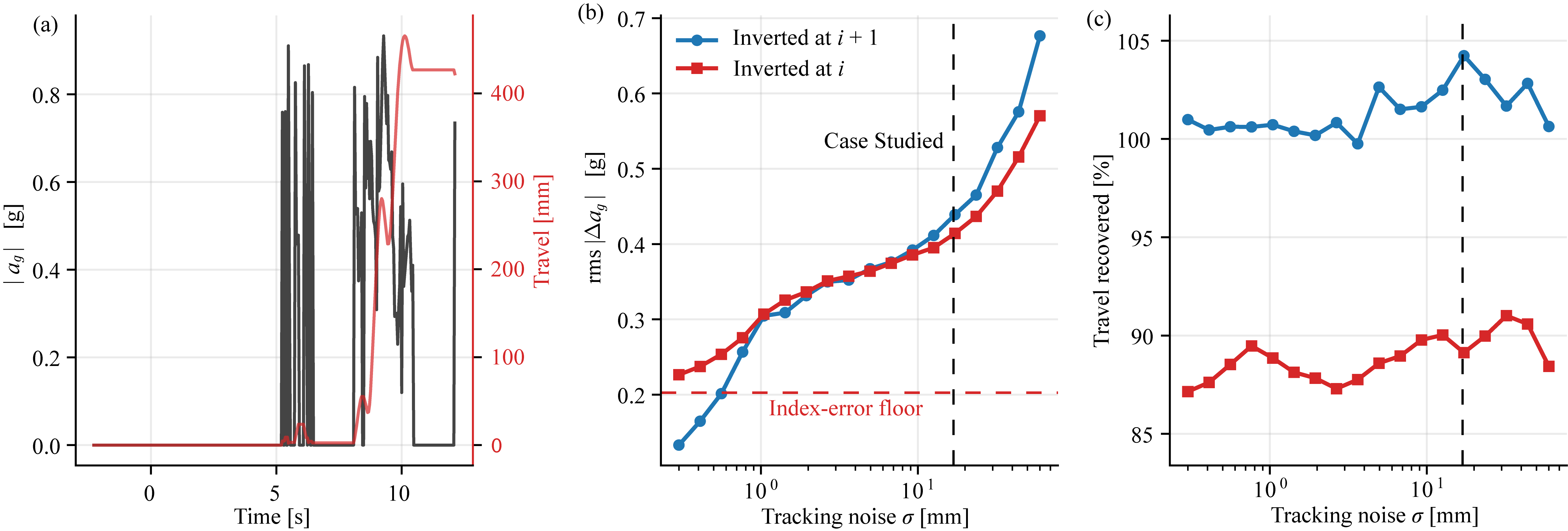}
  \caption{Reference solution built on measured kinematics, with synthetic tracking noise. (a) The reference: acceleration magnitude (dark, left axis) and cumulative travel (light, right axis). (b) Root-mean-square error of the recovered acceleration against noise level; the horizontal dashed line is the index-error floor of \SI{0.203}{\gravity} and the vertical dashed line is the noise level of the actual recording. Note that the two curves cross. (c) Travel recovered as a percentage of the reference; these curves do not cross at any noise level examined.}
  \label{fig:ruido}
\end{figure}

Zero-mean Gaussian noise of standard deviation $\sigma$ is added independently to each component of the tracked displacement, and $\sigma$ is swept over two orders of magnitude, from \SIrange{0.30}{60}{\milli\meter}, a range that brackets the noise level of the actual recording, $\sigma=\SI{16.9}{\milli\meter}$, by a factor of more than fifty below and more than three above. For each value of $\sigma$ and for each inverse the length of the smoothing window is re-optimized independently, so that neither scheme is handicapped by a filter that was chosen to suit the other. A window of length zero, meaning no smoothing at all, is included in the search, which is why the exact inverse returns exactly zero error at $\sigma=0$. \Cref{fig:ruido} reports the result and
\cref{tab:ruido} tabulates it.

\begin{table}[H]
  \centering
  \caption{Recovery from the measured-kinematics reference solution as a function of tracking noise. The root-mean-square columns cross at $\sigma\approx\SI{1.0}{\milli\meter}$, above which the inverse taken at step $i$ is the better of the two by that measure, whereas the travel columns do not cross at any noise level examined. The row at \SI{17.25}{\milli\meter} is the one nearest the noise floor of the recording, \SI{16.9}{\milli\meter}.}
  \label{tab:ruido}
  \begin{tabular}{rccccc}
    \toprule
    & \multicolumn{2}{c}{rms $\lvert\hat{\ag}-\ag\rvert$ [$g$]}
    & \multicolumn{2}{c}{travel recovered [\%]}\\
    \cmidrule(lr){2-3}\cmidrule(lr){4-5}
    $\sigma$ [mm] & at $i{+}1$ & at $i$ & at $i{+}1$ & at $i$\\
    \midrule
     0.00 & 0.0000 & 0.2025 & 100.0 & 85.5\\
     0.56 & 0.2013 & 0.2536 & 100.6 & 88.5\\
     1.43 & 0.3089 & 0.3255 & 100.4 & 88.1\\
     3.63 & 0.3522 & 0.3572 &  99.8 & 87.8\\
     9.25 & 0.3920 & 0.3856 & 101.6 & 89.8\\
    17.25 & 0.4390 & 0.4142 & 104.2 & 89.1\\
    23.56 & 0.4651 & 0.4367 & 103.0 & 90.0\\
    60.00 & 0.6765 & 0.5703 & 100.6 & 88.4\\
    \bottomrule
  \end{tabular}
\end{table}

Two readings of the same experiment point in opposite directions, and the disagreement between them is the substance of this subsection.

Judged by root-mean-square error the two inverses are comparable at the noise level of the actual recording, where the exact inverse gives \SI{0.439}{\gravity} and the inverse at $i$ gives \SI{0.414}{\gravity}, so that the incorrect one is in fact the lower of the two. The crossing occurs near $\sigma\approx\SI{1.0}{\milli\meter}$, which lies far below any realistic tracking floor for imagery of this kind. Taken in isolation, this panel would license the conclusion that the choice of index makes no difference in practice, or even that the inverse taken at step $i$ is to be preferred.

Judged by recovered travel the two are never comparable. The exact inverse returns between \SI{99.8}{\percent} and \SI{104.2}{\percent} of the reference travel across the entire sweep, whereas the inverse at $i$ returns between \SI{85.5}{\percent} and \SI{91.0}{\percent}, and the deficit is essentially flat in $\sigma$.

It should be pointed out that the quantity compared in this subsection is not the one compared in \cref{sec:costo}, and the two should not be read against each other. Here the recovered acceleration is compared against a reference acceleration that is known, which is possible only because the reference solution was manufactured, and the noise that degrades it is synthetic and is added on purpose. In \cref{sec:costo} no such reference exists, and what is compared is the relative acceleration that each inverse reproduces on re-integration against the relative acceleration measured in the field, with whatever noise the recording actually carries. The first comparison measures distance from the truth and the second measures internal consistency, and a scheme can rank differently on the two. The common conclusion, which is the one that matters, is that neither acceleration residual conveys the size or the direction of the error in the quantity being reported.

The explanation is that the two error sources are of different kinds. Tracking noise is zero mean, and the differentiation required to obtain acceleration from displacement amplifies it into a large-amplitude, high-frequency, sign-alternating contamination of the recovered acceleration, which inflates the root-mean-square error but integrates away. The index error is not zero mean. By \cref{eq:identidad} it equals $\mu_{k}g\cos\theta(\evk{i+1}-\evk{i})$, and for a sliding direction that turns consistently in one sense it points consistently in one sense, so that it survives two integrations and appears as a systematic shortfall in displacement. It is a bias, and a bias does not average out.

The methodological conclusion is general, and in our view it is insufficiently appreciated. For inverse problems in non-smooth dynamics, agreement in root-mean-square acceleration is not evidence that the inverse operator is the correct one. A scheme that is systematically wrong can rank ahead of the exact scheme on that metric once noise is present, because the metric rewards smoothness rather than consistency. Validation should instead be carried out on an integrated quantity such as recovered displacement, which is sensitive to bias and insensitive to zero-mean noise, and it should be carried out against a reference solution in which the answer is known by construction.

\section{Case Study: the 2026 La Guaira Earthquake}
\label{sec:aplicacion}

The preceding sections make a statement about an operator, and that statement stands independently of any particular data set. This section exercises the operator end to end on a real event, from an uncalibrated surveillance recording through to a peak ground acceleration with propagated uncertainty. We report the full chain rather than the final number alone, because in this class of problem the credibility of the estimate rests on the measurement chain that produced its input, and because the chain is itself a signal-processing problem of some
generality.

\subsection{Importance of the 2026 La Guaira Earthquake}
\label{sec:evento}

On 24 June 2026, two earthquakes of magnitude 7.2 and 7.5 occurred approximately \SI{39}{\second} apart on the Venezuelan Caribbean coast \citep{usgs2026laguaira}. Taken together, these earthquakes constitute the most destructive seismic disaster to affect Venezuela since the 1812 earthquake sequence. Although the 1900 (Mw 7.6) earthquake was the largest crustal earthquake documented in Venezuela during the instrumental era, its impacts were less severe than those associated with the 1812 events. The 2026 sequence produced widespread damage across north-central Venezuela, with the most severe impacts concentrated in La Guaira State \citep{yamazaki2005estimation,pousse2018earthquake,colon20151900,UNDRR2026Venezuela}. An event of this size in a densely built coastal corridor would ordinarily be expected to generate a substantial near-source data set against which ground motion models could be tested and against which the observed damage could be interpreted. In this case, it did not. Free-field strong-motion coverage of the affected region is sparse, the nearest instruments are distant from the districts that suffered the greatest damage, and the characterization of the shaking that actually occurred there consequently rests to an unusual degree on evidence that was never intended to serve as a measurement.

That circumstance is what gives the present case study its weight. In a well-instrumented region an estimate of peak ground acceleration derived from a surveillance video would be of methodological interest only, since an accelerograph would already have answered the question and answered it better. Here there is no such record to defer to. Any independent and quantitative constraint on the ground motion, provided its uncertainty is stated honestly and its measurement chain is documented well enough to be audited, is of direct value to the seismological and engineering reconstruction of the event, and observations that would ordinarily be treated as anecdotal acquire evidential standing. In that spirit, the recording analyzed below is treated as such, and the entire chain is reported rather than the final number alone.

A fixed surveillance camera recorded, at \SI{25}{\fps}, a
roadside scene containing several plastic delineator posts standing on a concrete surface. A second camera covered part of the same scene from a different vantage point; its recording is not analyzed here, because it shows more extensive pedestrian occlusion of the target post, and in the authors' judgment the first recording alone is sufficient to support the inversion. The sliding interface of interest is therefore the contact between the concrete and the delineator. During the strong-motion interval, several objects move significantly. However, the attention is placed on the rightmost delineator post, which slid visibly over the surface and came to rest permanently displaced. The Presidential Commission for the Assessment of Infrastructure Habitability in Venezuela provided the video.

Free-field strong-motion instrumentation is unevenly distributed, and in many regions of high seismic hazard the nearest recording station to a site of interest is ~12 to 20 kilometers away. Where no instrument was present, post-earthquake reconnaissance has long relied on displaced, overturned, or undisturbed objects left behind, which carry genuine information about the shaking that produced them. This reasoning is familiar in two established forms: the overturning of slender bodies, analyzed as a rocking problem since \citet{housner1963}, and the non-overturning of precariously balanced formations, which bounds the ground motion from above over geological time. Sliding bodies belong to the same family and have received less quantitative attention, although they are far more common, because a body that slides rather than topples records not a single threshold crossing but an entire history.

The contribution of an inversion such as \cref{eq:inverso} is to convert that qualitative practice into a measurement with a stated uncertainty. Two properties make it worth doing. The bound \cref{eq:cota} requires only the binary fact that the body moved, together with a coefficient of friction that can be measured afterward on the surviving hardware, so it applies even when no recording exists. And where a recording does exist, the inversion is exact and pointwise, so the recovered history inherits the resolution of the imagery rather than the resolution of a fitted source model. Surveillance cameras are dense in exactly the built environments where strong-motion instruments are sparse, which is the practical motivation for treating a recording of this kind as an instrument rather than as an anecdote.

That framing also sets the standard the measurement chain must meet. A surveillance recording is not an instrument: it is compressed, the camera mount itself moves because it is fixed to a building that vibrates during the event, the frame rate is nominally fixed and modest, the scene contains pedestrians who occlude the target, and nothing in it is calibrated. Each defect is addressed below, and each is accompanied by an independent check that does not use the information it tests.

Four stages separate the raw file from the input required by \cref{eq:inverso}. The camera motion is removed by registering each frame against static scene features, so apparent floor motion is not mistaken for post motion (\cref{sec:estabilizacion}). The imaging geometry is recovered from architectural constraints visible in the scene, which fixes a metric coordinate system on the ground plane and converts pixels to millimeters (\cref{sec:metrologia}). The post base is tracked through the sequence, including the intervals in which it is occluded, and the intervals in which the track is unreliable are identified from the correlation itself and bridged rather than inverted (\cref{sec:rastreo}). Finally the interface is characterized on the bench, which supplies the one parameter the imagery cannot
(\cref{sec:friccion}).

\subsection{Removal of Camera Motion}
\label{sec:estabilizacion}
\begin{figure}[htbp]
  \centering
  \includegraphics[width=0.5\linewidth]{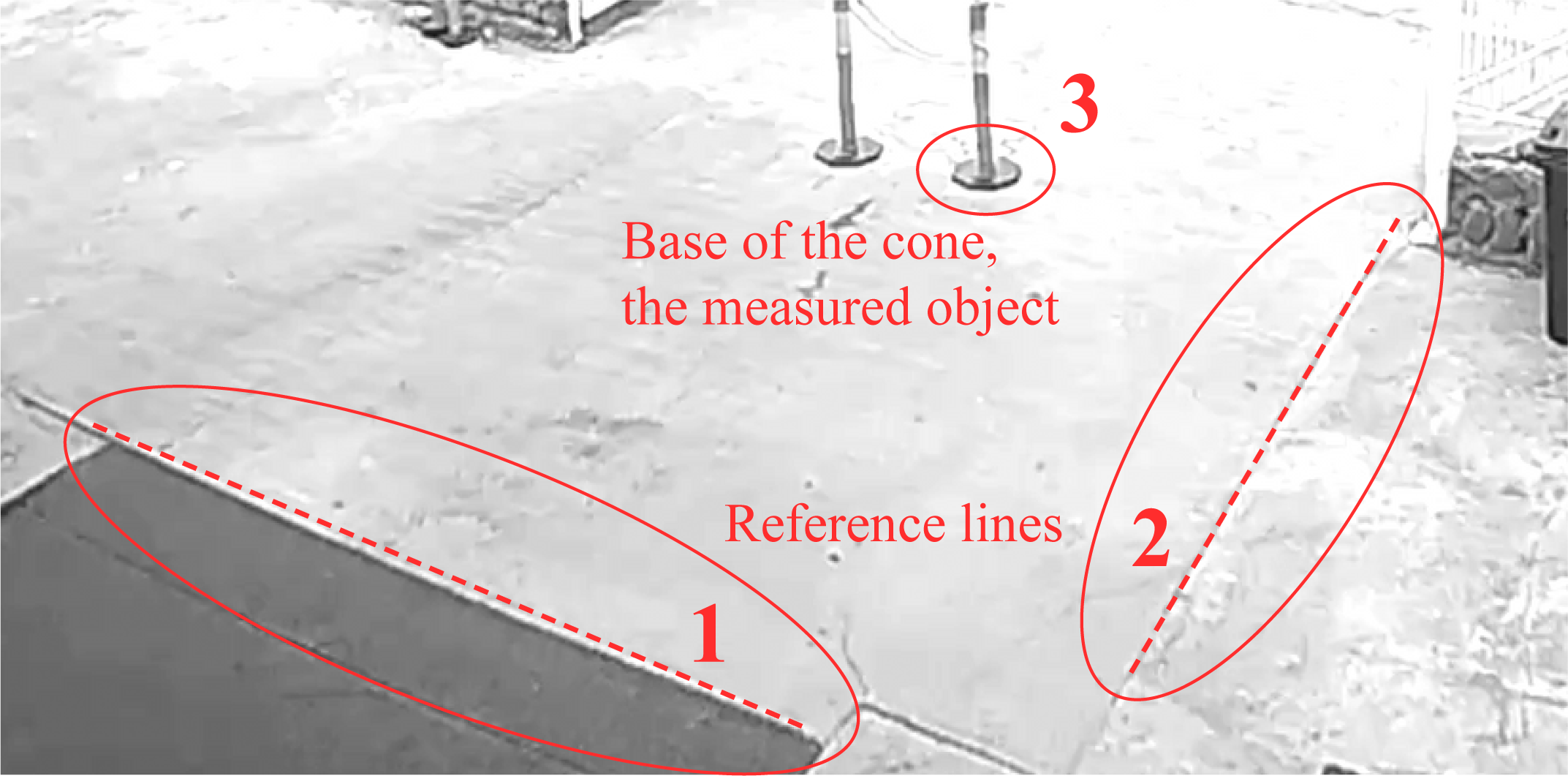}
  \caption{The two reference lines and the tracked object. Region 1 corresponds to the edge between the blue carpet and the floor. Region 2 corresponds to a floor joint. Region 3 is the base of the delineator post, which is the object measured.}
  \label{fig:referencias}
\end{figure}

The camera is rigidly mounted, but the mount vibrates during the event, so the image sequence shows apparent global motion that must be removed before measuring any relative displacement. The original recording runs 29 s; the analyzed clip begins \SI{2}{\second} into that recording, and every time reported below is measured from the start of the clip, not from the start of the original file. We perform this stabilization using the ground as the reference rather than any other moving object. The datum is the long straight edge where the floor meets the blue carpet, region 1 of \cref{fig:referencias}, some \SI{4.5}{\meter} from the post of interest. A straight edge is preferable to a patch on a pillar because it belongs unambiguously to the ground, because its position is overdetermined by tens of independent measurements along its length, and because its straightness in the ground plane is itself a check that can be carried out on the data. During the event this joint develops a visible crack. The crack falls in the dark, shadowed band of the image, outside the bright concrete band that the correlation tracks, so it does not enter the measured position.
 
The edge is fitted frame by frame from seventy-five profiles taken normal to it and correlated against the first frame, and the offsets along it are reduced to a straight line by robust least squares, whose intercept is the normal translation of the edge and whose slope is its apparent rotation. The fit cannot be used in pixels. The camera views the floor in strong foreshortening, and the image scale at the edge differs from that at the post, so the fitted edge is mapped point by point into the ground plane through the homography of \cref{sec:metrologia} and read in millimeters. The residual is \SI{0.00}{\milli\meter} over four meters of length, confirming that the feature is a physically straight line and not a mapping artifact. We then evaluate the rigid ground motion at the base of the post and return it to image coordinates through the local Jacobian. Over the sequence the edge translates through \SI{416}{\milli\meter} and rotates through \ang{2.8}, which is large compared with the slip to be measured and shows that the step is not optional.
 
A single straight line measures exactly two things, the rotation of the ground and the component of translation normal to itself. The parallel component is invisible, which is the aperture problem and not a defect of the fit, and a second line supplies it: a joint in the floor, region 2 of \cref{fig:referencias}, so the pair is well conditioned. The edge remains the datum, and the joint helps stabilize the component the edge cannot see.
 
\subsection{Scene Metrology}
\label{sec:metrologia}

Nothing in the recording is calibrated, so the imaging geometry must be recovered from the scene itself. Before any metric fit, every frame is corrected for lens (barrel) distortion with a Brown-Conrady radial-tangential model and a free principal point, fitted so that known straight features of the scene are returned straight; the fit reduces their root-mean-square deviation from straightness to \SI{0.70}{\pixel}, against several pixels under a purely radial model with the principal point fixed at the image center. All metrology below uses the corrected frames. The scene carries two families of mutually orthogonal straight features, the joints and edges of the floor and the lines of the fence footing, and the built environment supplies a third, vertical direction in the fence bars. These features are marked on the reference frame by hand, thirteen on the ground in the two horizontal directions and two on the pillars for the vertical, purely as an aid to identification; the markings are drawn on the digital image and are not present on the floor itself. Three orthogonal vanishing points determine the focal length and principal point in closed form \citep{caprile1990,hartley2004}, and the estimate is then refined under constraint using line evidence detected automatically in the scene \citep{gioi2010} rather than only the marked features. The refinement returns
\begin{equation}
  \bm{p}=(608.9,\;440.2)\ \si{\pixel},
  \qquad
  f=\SI{1108.4}{\pixel},
  \label{eq:calib}
\end{equation}
with a principal point substantially off center, which is expected of a cropped and rescaled surveillance stream and which the independent line evidence corroborates.

\begin{figure}[htbp]
  \centering
  \includegraphics[width=0.6\linewidth]{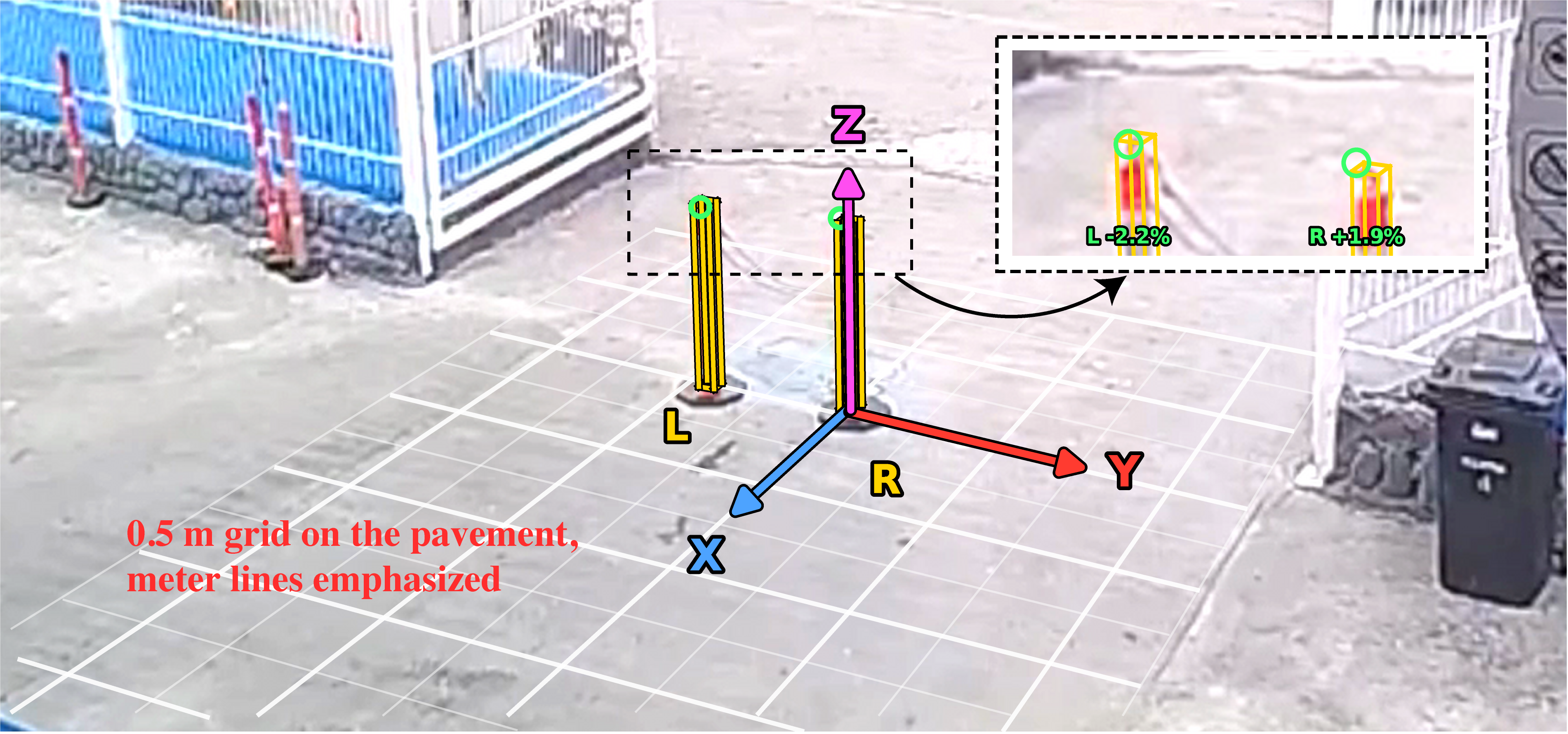}
  \caption{Verification of the recovered geometry on the region of interest of the annotated reference frame. A \SI{0.5}{\meter} metric grid generated from the ground-plane homography is projected onto the floor; the three global axes are drawn from the origin at the delineator base, and a box of height $h=\SI{1143}{\milli\meter}$ is erected on each measured post base (\textsf{R} right, \textsf{L} left). The boxes close on the observed tops of the posts, which is the visual form of the scale check.}
  \label{fig:grid}
\end{figure}

\begin{figure}[htbp]
  \centering
  \includegraphics[width=0.38\linewidth]{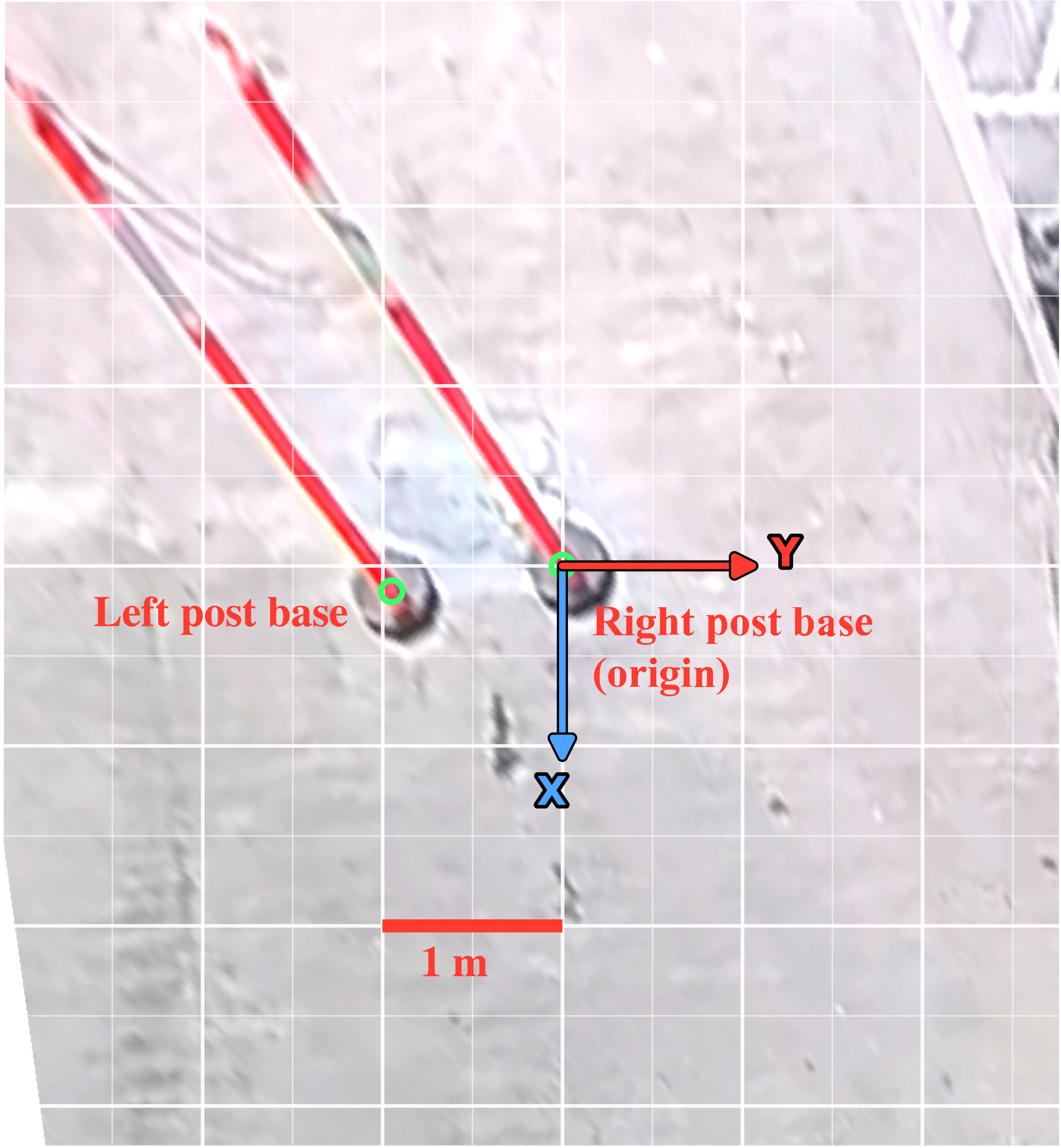}
  \caption{Orthographic top view obtained by resampling the reference frame through the inverse homography at \SI{6}{\milli\meter} per pixel. Grid spacing is \SI{0.5}{\meter}, with meter lines emphasized. The origin is the base of the rightmost delineator. Objects above the floor are smeared radially, as expected from ground-plane rectification.}
  \label{fig:top}
\end{figure}

The ground-plane homography follows, and its metric scale is fixed by the known height of the delineator posts, following the single-view metrology of \citet{criminisi2000}. The two posts stand at different depths and subtend different numbers of pixels, yet they return scale factors agreeing to \SI{2.9}{\percent}; a calibration error in $f$ or in $\bm{p}$ would produce a depth-dependent bias and would break that agreement, so the check is informative. \Cref{fig:grid} shows the recovered geometry projected back onto the reference frame, and \cref{fig:top} the orthographic rectification of the floor.

The uncertainty of the recovered geometry is dominated by choices rather than by pixel noise, i.e., which is the top of a post, and what the post height actually is. Forcing the principal point to the image center displaces the result by more than four times the overall spread, which is what justifies recovering it in \cref{eq:calib} rather than assuming it. And the metric scale enters linearly, so an error of \SI{5}{\percent} in the assumed post height propagates to \SI{5}{\percent} in every distance reported here, and a reader who doubts that height can rescale the results accordingly.

\subsection{Handling Occlusions and Measuring Slip}
\label{sec:rastreo}

The post base is located in every frame by normalized cross-correlation against a template, using features selected as in \citet{shi1994}. Because pedestrians cross the target during the event, a greedy frame-to-frame tracker fails; i.e., at an occluded frame there is no information at that frame on which to base a decision. The correlation surfaces are therefore stacked over the whole sequence, and the trajectory is chosen by dynamic programming as the globally optimal path under a smoothness penalty \citep{viterbi1967,felzenszwalb2011}, so that the frames in which the target is visible and the correlation is sharp dictate the path through the frames in which it is not. Of the \num{361} frames, \num{319} are classified as good, \num{13} marginal, and \num{29} rejected, with the rejections forming five contiguous intervals corresponding to people passing across the target. The displacements are referred to a coordinate system attached to the ground surface in the scene rather than to the image frame, so that whatever registration error survives the stabilization of \cref{sec:estabilizacion} does not enter the measurement as apparent motion of the post.

The template match returns a normalized correlation coefficient for every frame, and the frames in which that coefficient collapses are the frames in which a body crosses in front of the post. Outside the occlusions and before the onset of motion, the coefficient averages \num{0.958}; on the occluded frames, it averages \num{0.272}. A gate at \num{0.45} flags \num{28} frames, and closing the single-frame gaps between them brings the count to \num{29}, falling in five intervals: \SIrange{7.04}{7.12}{\second}, \SIrange{7.24}{7.68}{\second}, the single frame at \SI{7.84}{\second}, \SIrange{9.56}{9.88}{\second} and \SIrange{10.56}{10.68}{\second}. A further \num{13} frames lie between \num{0.45} and \num{0.70} and are carried as marginal. Across each interval, linear interpolation between the reliable frames on either side bridges the displacement, so neither tracking failure nor the pedestrian impact enters the inversion as an acceleration.

\begin{figure}[htbp]
  \centering
  \includegraphics[width=0.7\linewidth]{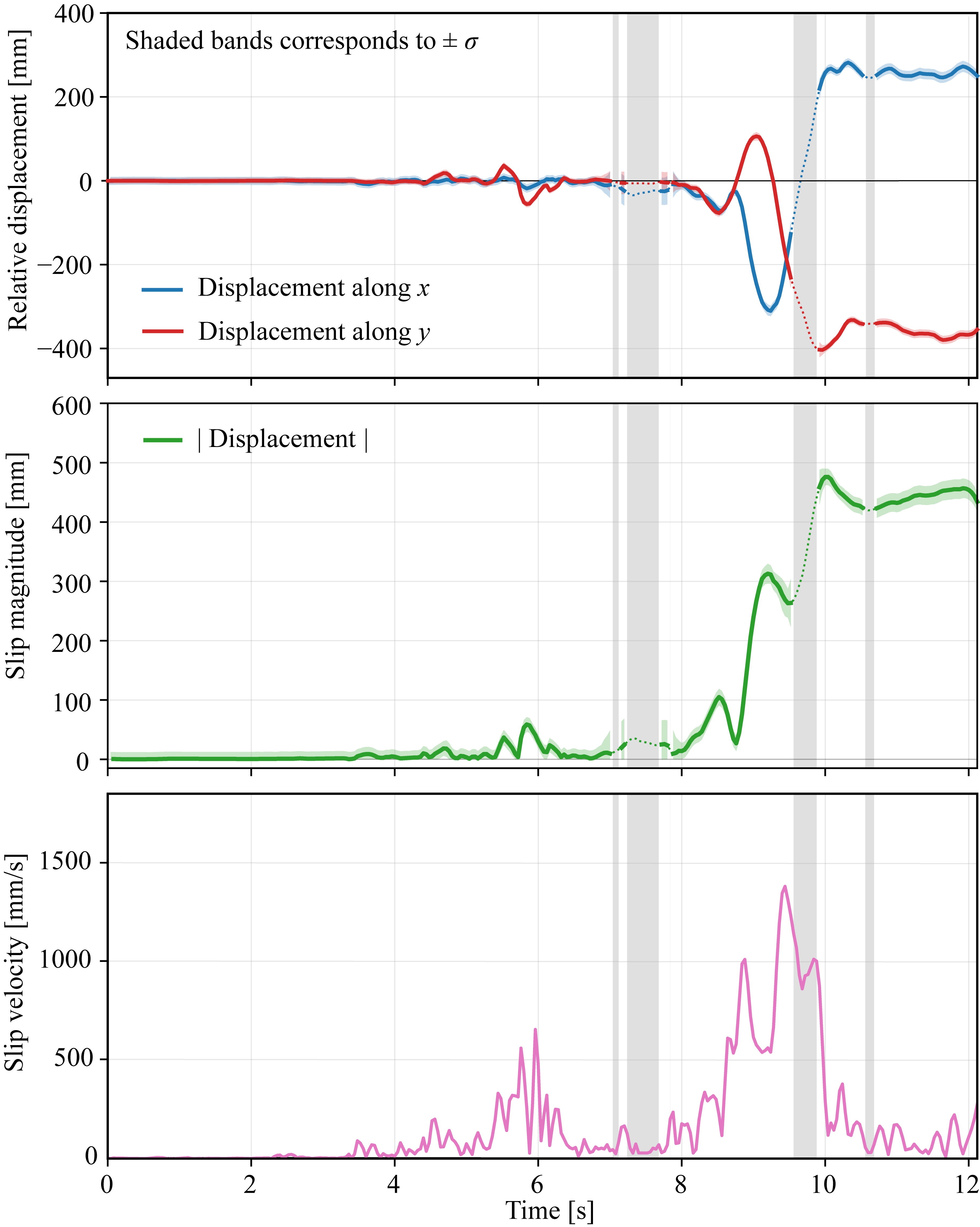}
  \caption{Sliding of the right delineator post relative to the ground.  Top, the two components with their standard-uncertainty bands. Middle, the slip magnitude. Bottom, the slip rate. Gray bands and dotted lines mark the five intervals rejected for occlusion and filled by interpolation.}
  \label{fig:slip}
\end{figure}

\begin{figure}[htbp]
  \centering
  \includegraphics[width=0.7\linewidth]{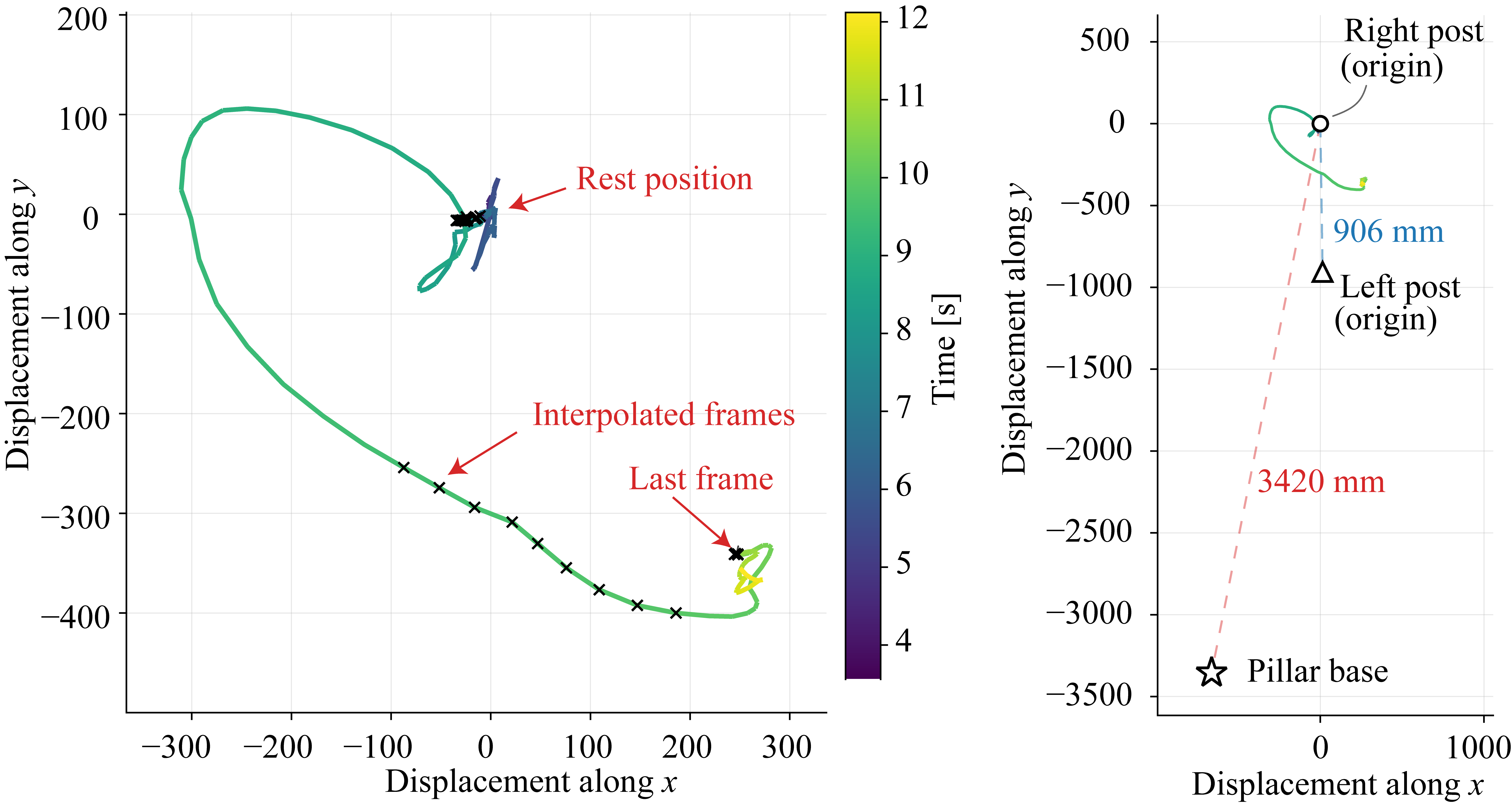}
  \caption{Left, the trajectory of the base in the ground plane, colored by time, the color scale shows the interval in which the base moves rather than over the whole record. Right, a location map at true scale carrying the pillar base, the left post, and the post of interest.}
  \label{fig:traj}
\end{figure}

The peak recovered in \cref{sec:recuperado} falls at
$t=\SI{9.24}{\second}$ and the peak slip rate at $t=\SI{9.52}{\second}$, both on measured frames and both outside all five intervals, the nearest of which begins at \SI{9.56}{\second}. \Cref{fig:slip} presents the resulting slip history and \cref{fig:traj} the trajectory in the ground plane; \cref{tab:rastreo} collects the principal figures.

\begin{table}[htbp]
  \centering
  \caption{Recovered sliding of the right delineator post, measured against the static ground datum of \cref{sec:estabilizacion}. $x$ and $y$ are the two horizontal directions of \cref{eq:calib}, marked in blue and in red respectively on \cref{fig:grid}. The peak slip rate is read from the smoothed track that the inversion uses as input.}
  \label{tab:rastreo}
  \begin{tabular}{lr}
    \toprule
    quantity & value\\
    \midrule
    net displacement $\Delta x$       & \SI{+257}{\milli\meter}\\
    net displacement $\Delta y$       & \SI{-362}{\milli\meter}\\
    net slip magnitude                & \SI{444}{\milli\meter}\\
    direction relative to $x$         & \ang{-54.6}\\
    standard deviation over the window & \SI{16}{\milli\meter}\\
    standard uncertainty, final frame & $13$~\si{\milli\meter}\\
    peak slip rate (measured frames)  &
      \SI{1179}{\milli\meter\per\second} at \SI{9.52}{\second}\\
    onset of sustained motion         & \SI{8.88}{\second}\\
    \bottomrule
  \end{tabular}
\end{table}

Three features of the history bear directly on the inversion. First, the post rides with the floor for a long time before it gives way. The ground datum itself begins to move at $t=\SI{4.40}{\second}$, while the slip stays below \SI{75}{\milli\meter} until $t=\SI{8.40}{\second}$ and sustained motion is declared at $t=\SI{8.88}{\second}$. The post therefore rides for some \SI{4}{\second} before friction is overcome, and by \cref{eq:stick} that interval constrains the excitation.

The second is that the slip is not a single push. The smoothed magnitude passes \SI{142}{\milli\meter} at \SI{8.88}{\second}, reaches \SI{294}{\milli\meter} at \SI{9.20}{\second}, falls back to \SI{250}{\milli\meter} at \SI{9.48}{\second} and then climbs to \SI{467}{\milli\meter} at \SI{10.08}{\second}. Between \SI{8.40}{\second} and \SI{10.80}{\second} the base travels a path of total length \SI{1192}{\milli\meter} to achieve a net displacement of \SI{413}{\milli\meter} over the same interval, a ratio of \num{2.9}. The net slip is the residue of a reversing motion and not the integral of a one-directional one.

The third is that the reversal changes direction, not merely sign along a line. The base first moves out to \SI{294}{\milli\meter} in a direction of \ang{173.1}, almost opposite to where it ends, and then swings through \ang{133} to finish at \ang{-54.6}; the $x$ coordinate alone covers \SIrange{-292}{280}{\milli\meter}. A one-dimensional friction model cannot produce this, because the direction of the friction force is not free once the direction of sliding is fixed. This feature makes the record a two-dimensional problem and therefore one on which the identity \cref{eq:identidad} is active.

\begin{figure}[htbp]
  \centering
  \includegraphics[width=0.72\linewidth]{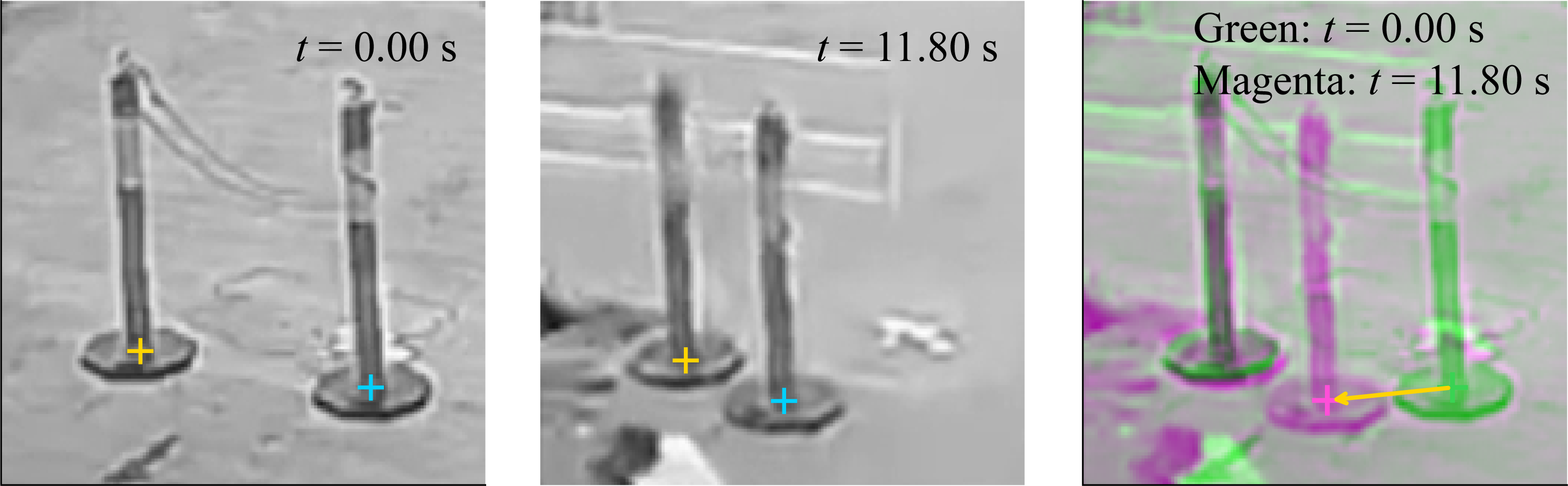}
  \caption{Frames corresponding to $t=0$s and $t=11.80$s and their superposition. The left post coincides with itself to \SI{5.4}{\milli\meter}; the post of interest is displaced by \SI{455}{\milli\meter}. The comparison frame is the one in the rest window that maximizes the lesser of the two template correlations, rather than one
  chosen by hand.}
  \label{fig:beforeafter}
\end{figure}

A direct verification independent of the tracker is available and is shown in \cref{fig:beforeafter}: two frames of the region of interest, aligned and superimposed. The alignment uses no tracking information. The left post is not static during the event, i.e., it moves substantially and is struck by a pedestrian near $t=\SI{9.7}{\second}$. What the comparison shows is that its initial and final positions nearly coincide, by \SI{5.4}{\milli\meter}, which is where the collision happened to leave it rather than evidence that the post stood still. So the left post serves here as a static control. The control reads \SI{5.4}{\milli\meter} while the post of interest is displaced by \SI{455}{\milli\meter}, a ratio of better than eighty to one.

The smoothed displacement used as input to the inversion has an endpoint of \SI{444.2}{\milli\meter} over the rest window, in agreement with the raw \SI{444.5}{\milli\meter} of \cref{tab:rastreo} to three tenths of a millimeter, and a residual noise floor of \SI{16.9}{\milli\meter}. Stationary intervals are declared below a slip rate of \SI{109}{\milli\meter\per\second}, which is three standard deviations of the velocity over the pre-event interval.

\subsection{Characterization of the Friction Coefficients}
\label{sec:friccion}

The friction coefficients are the one input the imagery cannot supply, and they dominate the final uncertainty. They were measured afterward on the surviving hardware, on the posts themselves and on the floor material, with a hand dynamometer applied horizontally at the base of the post so that no overturning moment is generated \citep{boroschek2026bowling}. Two quantities have to be kept apart. The static coefficient $\mu_{s}$ sets the stick condition of \cref{eq:moreau} and therefore decides when motion begins; the kinetic coefficient $\mu_{k}$ is the one that enters the inverse of \cref{eq:inverso} and sets the scale of the recovered acceleration. Both were obtained from the same tests, read in two different ways. With $mg=\SI{56.29}{\newton}$, the forces convert directly to coefficients.

Twenty-six breakaway readings were taken, thirteen pushing and thirteen pulling, on two posts, and on the second post along two different directions. Each is the maximum force reached immediately before the post began to move, so each measures $\mu_{s}$ and not $\mu_{k}$.

The data show a systematic difference between the two protocols: pushing averages \SI{37.56}{\newton} and pulling \SI{32.63}{\newton}, a ratio of \num{1.151} with $p=\num{1.2e-4}$ on a two-sample $t$ test. Since the force is applied at the base this cannot be an overturning moment, and it is not a property of the interface. It is the signature of a dynamometer held slightly off horizontal: if the applied force makes an angle $\alpha$ with the floor, its normal component adds to the weight when pushing and subtracts when pulling, so that
\begin{equation}
  F_{\mathrm{push}}=\frac{\mu\,mg}{\cos\alpha-\mu\sin\alpha},
  \qquad
  F_{\mathrm{pull}}=\frac{\mu\,mg}{\cos\alpha+\mu\sin\alpha}.
  \label{eq:pushpull}
\end{equation}
The arithmetic mean of the two is biased at first order in $\alpha$; their product is not, because the cross terms cancel. The unbiased estimator is therefore the geometric mean,
\begin{equation}
  \hat{\mu}
  =\frac{\sqrt{F_{\mathrm{push}}F_{\mathrm{pull}}}}{mg}
  =\frac{\mu}
        {\sqrt{\cos^{2}\alpha-\mu^{2}\sin^{2}\alpha}},
  \label{eq:mugeom}
\end{equation}
and the two force columns are multiplied rather than averaged. Inverting the observed ratio gives $\alpha=\ang{6.5}$, a plausible hand-held tilt, at which the residual bias in \cref{eq:mugeom} is \SI{0.9}{\percent} and can be neglected. We record this because the correction is worth about \SI{7}{\percent} in the coefficient and therefore, by the proportionality established in \cref{sec:barridos}, about the same in the recovered peak.

\begin{figure}[htbp]
  \centering
  \includegraphics[width=0.6\linewidth]{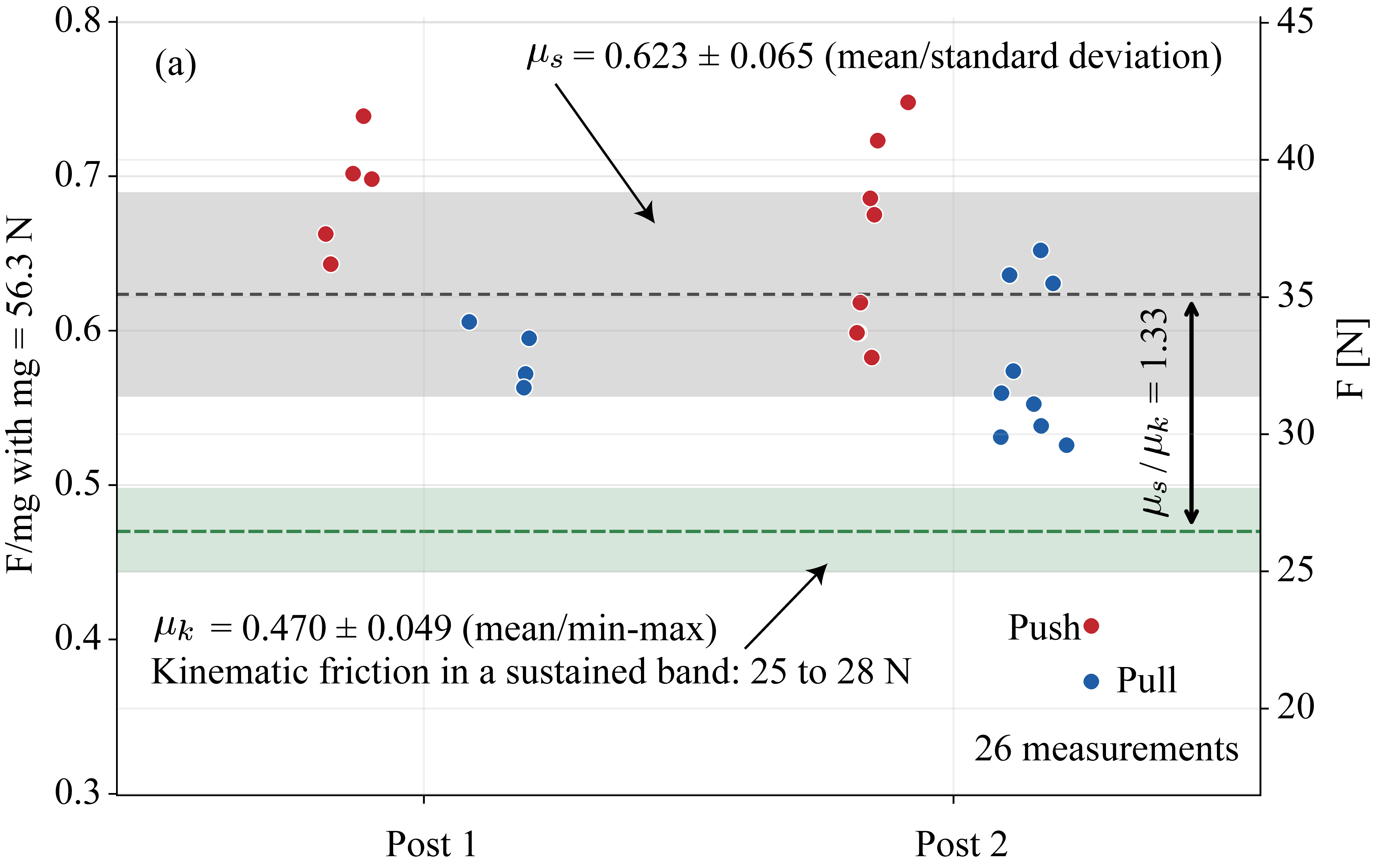}
  \caption{The twenty-six dynamometer readings, which are breakaway maxima and therefore estimate $\mu_{s}$ of \cref{eq:mus}, with the separation between the push and pull protocols (15\%) that motivates the geometric-mean estimator of \cref{eq:mugeom}, and beneath them the band of sustained force that supplies the $\mu_{k}$ of \cref{eq:muk}.}
  \label{fig:friction}
\end{figure}

Pooling all twenty-six readings, the static friction coefficient is described as
\begin{equation}
  \mu_{s}=\num{0.6235}\pm\num{0.0655},
  \label{eq:mus}
\end{equation}
where the quoted dispersion is the sample standard deviation of the individual readings and therefore includes repeatability, the residual push and pull bias, and any real directional dependence. A Shapiro-Wilk test \citep{shapiro1965} returns $W=\num{0.958}$ with $p=\num{0.35}$, so a normal model for the sampling distribution is not contradicted and is adopted later in \cref{sec:propagacion}. \Cref{fig:friction} shows the readings.

The kinetic coefficient was taken from the same tests. Once the post is sliding, the reader does not settle on a single value but oscillates between \SI{25}{\newton} and \SI{28}{\newton}, and it does so under both protocols, so the pair is a band of sustained force rather than a push and pull pair, and \cref{eq:mugeom} does not apply to it as such. The distinction turns out to be immaterial, since the arithmetic and geometric centers of the band differ by \SI{0.16}{\percent}. Its width is \SI{11.3}{\percent} in relative terms, which is smaller than the \SI{15.1}{\percent} push and pull asymmetry of the breakaway readings, so the oscillation is consistent with being the same hand tilt seen again rather than a property of the interface. Taking the center of the band,
\begin{equation}
  \mu_{k}=\num{0.470}\pm\num{0.052}.
  \label{eq:muk}
\end{equation}
The dispersion in \cref{eq:muk} is not the band width. Read alone, a uniform force within the band would give a coefficient of variation of only \SI{3.3}{\percent}, which measures how well the needle can be read and nothing else. It says nothing about the fact that the coefficient varies from point to point on the slab and from post to post, and that variation is precisely what the twenty-six breakaway readings measure, at \SI{10.5}{\percent}. The two contributions are independent and are combined, the second dominating. The resulting ratio $\mu_{s}/\mu_{k}=\num{1.33}$ is unremarkable for a polymer sliding on concrete and satisfies the requirement $\mu_{s}\ge\mu_{k}$ imposed in
\cref{sec:formulacion}.

\subsection{Survey of the Slab}
\label{sec:pendiente}

The inclination of the concrete surface and the direction of its steepest descent enter \cref{eq:inverso} as the gravity term. Neither can be read from the imagery, so a team of researchers from the Civil Engineering Department at the University of Chile measured both directly on the floor in a field survey carried out after the earthquake \citep{boroschek2026bowling}. They used a digital inclinometer over the floor region traversed by the post. Forty-two stations were occupied, arranged in six blocks of seven, and at each station the instrument was read three times along each of the two horizontal axes of the global frame of \cref{sec:metrologia}. The two dispersions in the resulting set are different and worth separating. Repeat readings at a fixed station scatter by \ang{0.19} along the first axis and \ang{0.26} along the second, which is the instrument's repeatability. Station means scatter by \ang{1.09}, which is five times larger and is not measurement error: the slab is genuinely uneven at the meter scale. Taking the resultant $\theta=(\theta_{x}^{2}+\theta_{y}^{2})^{1/2}$ station by station and discarding one station that lies \num{4.5} robust standard deviations above the rest gives
\begin{equation}
  \theta=\ang{2.38}\pm\ang{0.92},
  \label{eq:theta}
\end{equation}
over forty-one stations. The sample is well described by a Gaussian. Against a lognormal, which would be the natural alternative for a positive quantity, the Gaussian wins on all three tests we applied: Kolmogorov-Smirnov $D=\num{0.094}$ against \num{0.112}, Shapiro-Wilk $W=\num{0.965}$ at $p=\num{0.23}$ against \num{0.947} at $p=\num{0.06}$, and Anderson-Darling $A^{2}=\num{0.39}$ against \num{0.64} for a \SI{5}{\percent} critical value of \num{0.73}. Only \SI{0.5}{\percent} of the Gaussian mass lies below zero, and it is removed by rejection when the distribution is sampled in \cref{sec:propagacion}.

The direction of steepest descent follows from the signs of the two components, which the inclinometer records along with their magnitudes. It runs toward negative $x$ and positive $y$ in the frame of \cref{sec:metrologia}, that is at an azimuth of \ang{161.8}. The post did not slide in that direction. Its net displacement lies at \ang{-54.6}, which is \ang{144} away from the downslope azimuth and within \ang{37} of straight up the slope. This does not conflict with the mechanics. The slope term contributes $g\sin\ang{2.38}=\SI{0.042}{\gravity}$, which is negligible compared with the excitation recovered below, and the direction a block ends up after a strong two-dimensional shake is set by the excitation, not by a gravity term twenty times smaller. The survey is therefore doing real work here: the azimuth is an independent input to \cref{eq:inverso} and cannot be inferred from the trajectory it helps explain.

\subsection{Recovered Ground Acceleration}
\label{sec:recuperado}

\begin{figure}[htbp]
  \centering
  \includegraphics[width=\linewidth]{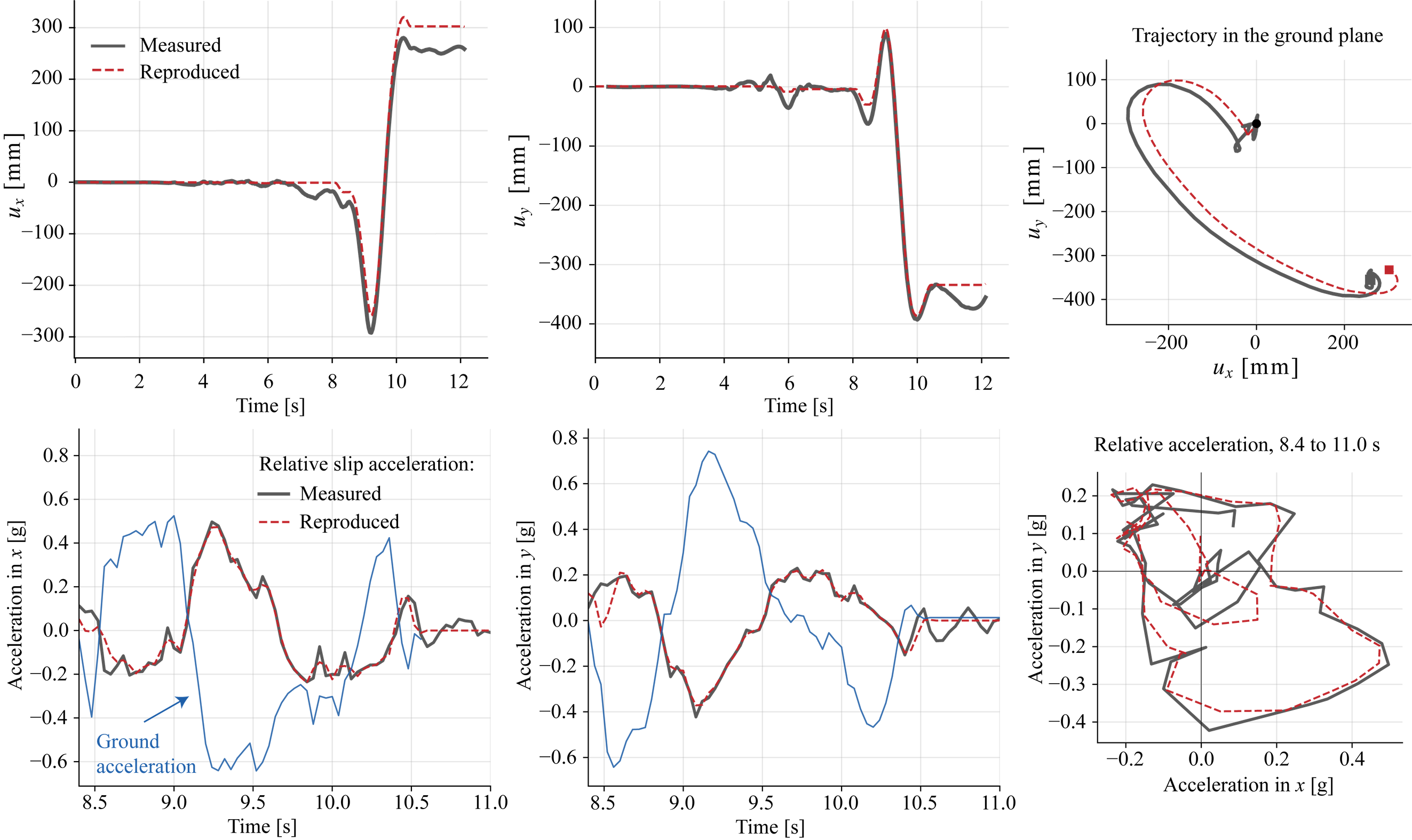}
  \caption{Forward verification at the measured $\mu_{k}=\num{0.47}$. The measured slip against the motion produced by integrating \cref{eq:moreau} with the recovered ground acceleration, shown as both components against time and as the trajectory in the ground plane. The upper panels show measured traces through the \SI{1.00}{\second} displacement filter. The lower panels use the \SI{0.20}{\second} acceleration filter and show three distinct signals: the measured relative acceleration of the post, the same quantity reproduced by the forward simulation, and the recovered ground acceleration.}
  \label{fig:forward}
\end{figure}

Two smoothing operations enter the pipeline, they act on different signals, and separating them is a precondition for reading the results that follow. The first is applied to the tracked displacement, before any inversion takes place: a third-order Savitzky-Golay filter over twenty-five frames, which at the \SI{25}{\hertz} frame rate is a window of \SI{1.00}{\second}. Its purpose is to make the second difference of a signal quantized to the pixel usable at all, and it is this smoothed displacement, not the raw track, that the velocity threshold of \cref{sec:rastreo} sorts into sliding and stationary frames. The second is applied to the ground acceleration that \cref{eq:inverso} returns: a third-order Savitzky-Golay filter over five frames, or \SI{0.20}{\second}. Its purpose is to suppress the frame-to-frame scatter that survives differencing twice, and it is the window quoted in the robustness sweep below and in the caption of \cref{fig:forward}. Neither filter touches the friction law or the time stepper. Both act on sampled signals, one before the inversion and one after it, and no property of \cref{prop:inverso} depends on either.

Applying \cref{eq:inverso} frame by frame over the sliding steps returns the ground acceleration directly, with no regularization and no fitted source model. At the central values of \cref{eq:muk,eq:theta} and the surveyed azimuth the peak recovered magnitude is \SI{0.912}{\gravity}, reached at $t=\SI{9.24}{\second}$ in a direction of \ang{133.4}. It falls inside the analysis window rather than at its edge, and on a measured rather than an interpolated frame, so it is a resolved feature of the record and not an artifact of where the window was cut or of where the occluded frames were bridged.

Two checks bound the credibility of that number before propagating any uncertainty. Re-integrating \cref{eq:moreau} with the recovered sequence, taken unsmoothed, returns the velocity sequence the scheme propagates to \SI{4.8e-13}{\milli\meter\per\second} and the corresponding trajectory to \SI{2.1e-13}{\milli\meter}, which is machine precision and confirms that the implementation realizes \cref{prop:inverso} rather than an approximation of it. Separately, forward integration of the \emph{smoothed} accelerogram reproduces the measured trajectory with a root-mean-square departure of \SI{30.0}{\milli\meter} against a measurement noise floor of \SI{16.9}{\milli\meter}, so that the reproduction is good to about one and three-quarters times the noise, the residual departure being the cost of the smoothing rather than of the inversion. The net displacement recovered is \SI{448.9}{\milli\meter} against a measured \SI{428.9}{\milli\meter}, an overshoot of \SI{4.7}{\percent}. \Cref{fig:forward} shows the comparison.

The peak is stable across smoothing-window choices. Widening the acceleration filter from \SIrange{0.20}{0.52}{\second} lowers the peak monotonically by \SI{5.4}{\percent}, and removing it altogether raises the peak only to \SI{0.933}{\gravity}, \SI{2.3}{\percent} above the reported value.

The natural next question is whether the recovered accelerogram also delivers a peak ground velocity, which is the quantity most directly tied to structural damage. It does not. The ground acceleration is determined only while the post slides, which is seventy-two of the three hundred and sixty-one frames. Over the remaining two hundred and eighty-nine, the equation of motion returns nothing about the ground beyond an inequality, i.e., the acceleration stayed inside the sticking disc $\lVert\ag-g\sin\theta\,\ed\rVert\leq\mu_{s}g\cos\theta$, whose radius of \SI{0.62}{\gravity} is two thirds of the peak itself. What \cref{eq:inverso} emits on those frames is the single value $g\sin\theta\,\ed$, with magnitude \SI{0.0416}{\gravity}; it is the center of that disc, not a measurement of anything. It should be read as an absence of data rather than as a quiet ground, and over four-fifths of the record is what the video contains.

The recovered peak ground velocity depends on the high-pass filter chosen before integration. Figure 12 shows that, across a reasonable range of cut-off frequencies, the resulting curves agree on the time of the peak, $t=\SI{9.08}{\second}$, but not on its value, which ranges from \SI{897}{\milli\meter\per\second} to \SI{1960}{\milli\meter\per\second}, a factor of \num{2.19}. The result is further complicated by the window used to recover the slip, which appears to include an interval from about \SI{9.40}{\second} to \SI{9.56}{\second}, in which a pedestrian may have been in contact with the post; if so, the motion recovered over that interval would not be attributable to the ground alone. For both reasons, we do not report a peak ground velocity.

\begin{figure}[htbp]
  \centering
  \includegraphics[width=\linewidth]{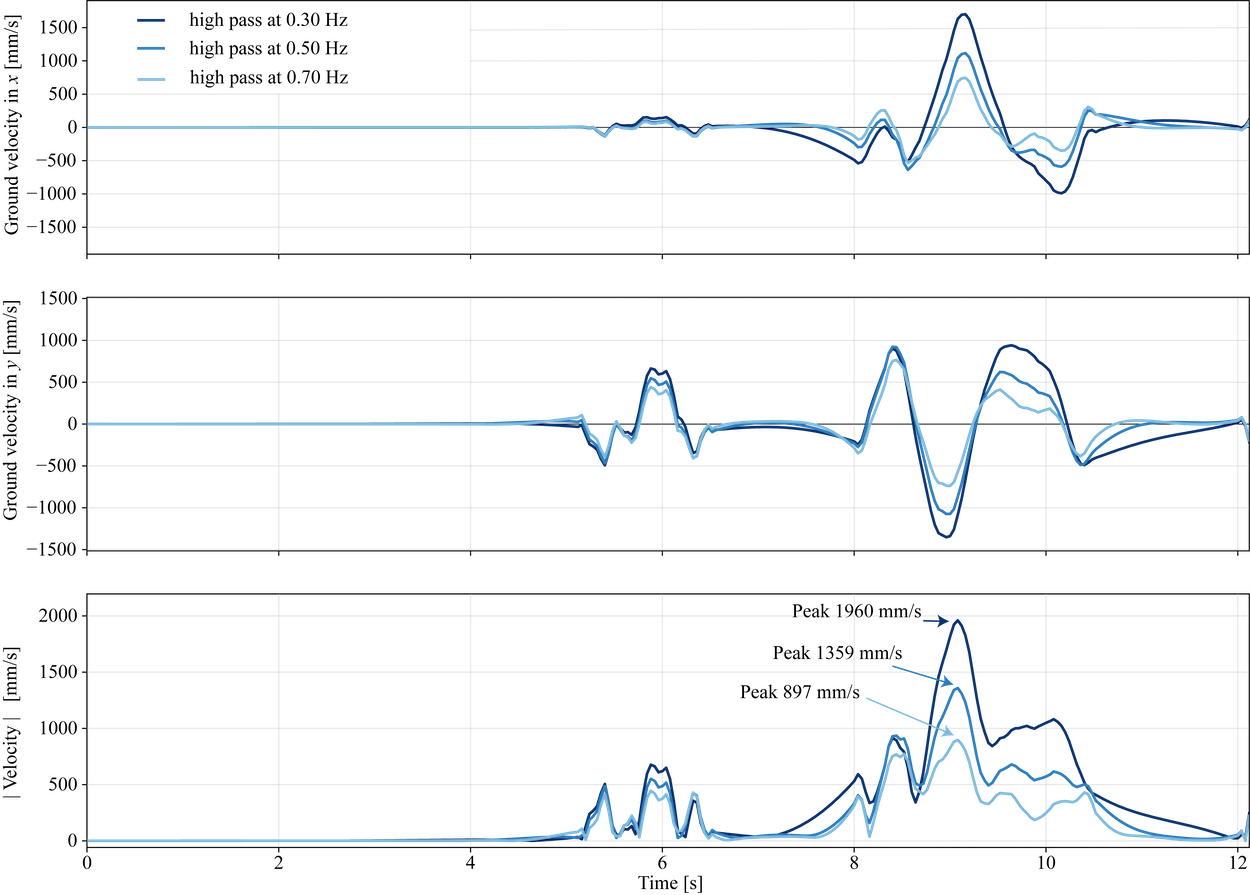}
  \caption{Plots (a) and (b) give the two components of the integrated velocity and panel (c) its magnitude, each for three high-pass cut-offs that an analyst could equally defend. The three agree on the peak time but differ in its value by a factor of \num{2.19}; there is no plateau, which is the signature a resolved velocity would leave.}
  \label{fig:determinada}
\end{figure}


\subsection{Uncertainty Propagation}
\label{sec:propagacion}

\begin{figure}[htbp]
  \centering
  \includegraphics[width=\linewidth]{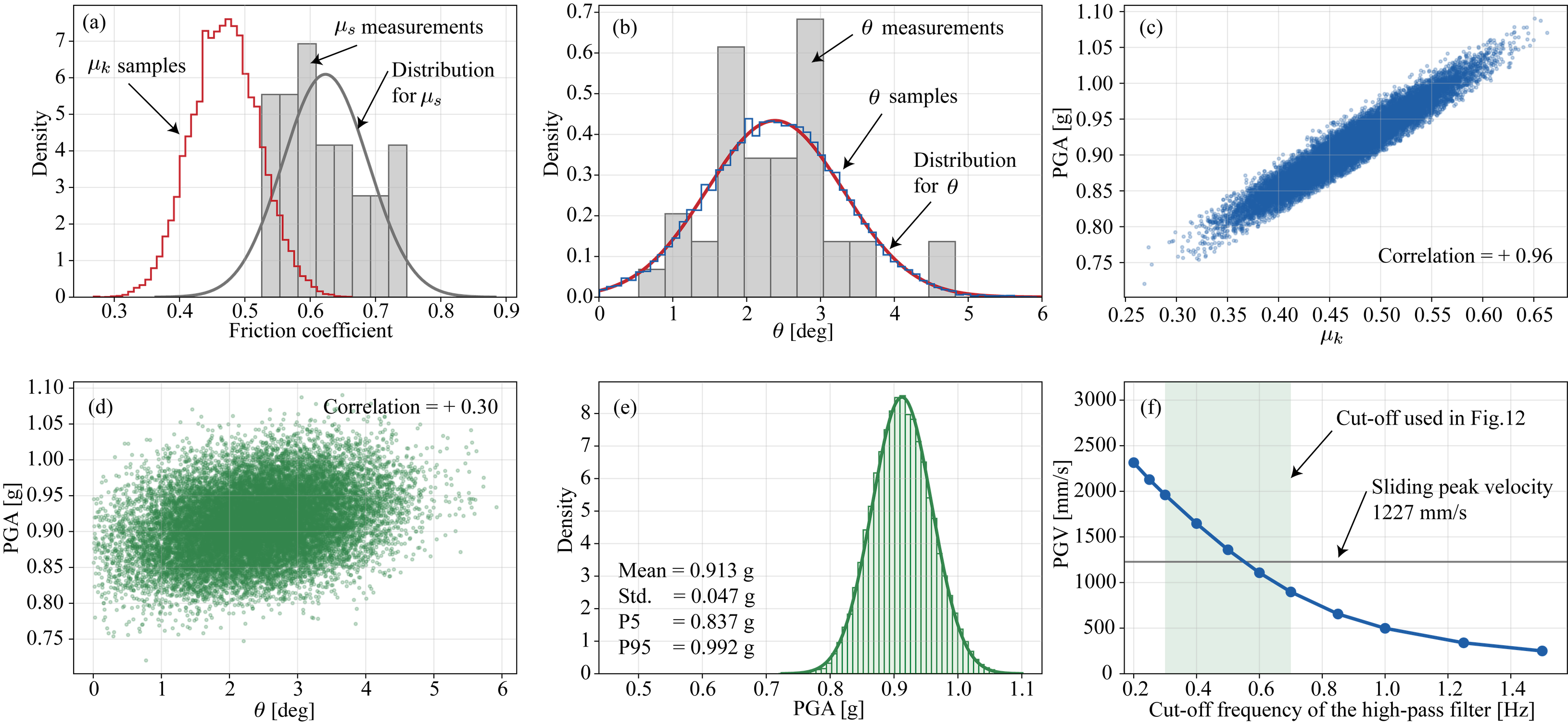}
  \caption{Monte Carlo propagation, \num{20000} samples. (a) Probabilistic model adopted for $\mu_s$ and $\mu_k$ using Gaussian distributions. (b) Probabilistic model adopted for $\theta$, corresponding to a truncated Gaussian distribution. (c) Samples of PGA against $\mu_k$ and its correlation. (d) Samples of PGA against $\theta$ and its correlation. (e) Histogram for the PGA identified. (f) Peak ground velocity against the cut-off frequency of the high-pass filter used before integrating.}
  \label{fig:mc}
\end{figure}

Monte Carlo propagates the two parameters the kinematics cannot determine. The kinetic coefficient is drawn from \cref{eq:muk}, as a uniform draw within the observed force band multiplied by the interface variability of \cref{eq:mus}; the inclination is drawn from the Gaussian fitted to the forty-one surveyed stations in \cref{eq:theta}, truncated at zero by rejection rather than by clipping, since clipping would pile mass at the origin and bias the mean. The azimuth is held at its surveyed value. Twenty thousand samples give
\begin{equation}
  \mathrm{PGA}=\num{0.912}\pm\SI{0.047}{\gravity},
  \label{eq:pga}
\end{equation}
a coefficient of variation of \SI{5.1}{\percent}, with a median of \SI{0.912}{\gravity} and a fifth to ninety-fifth percentile range of \SIrange{0.837}{0.992}{\gravity}. \Cref{fig:mc} shows the inputs, the
dependence of the peak on each of them, and the output distributions.

\begin{figure}[htbp]
  \centering
  \includegraphics[width=0.6\linewidth]{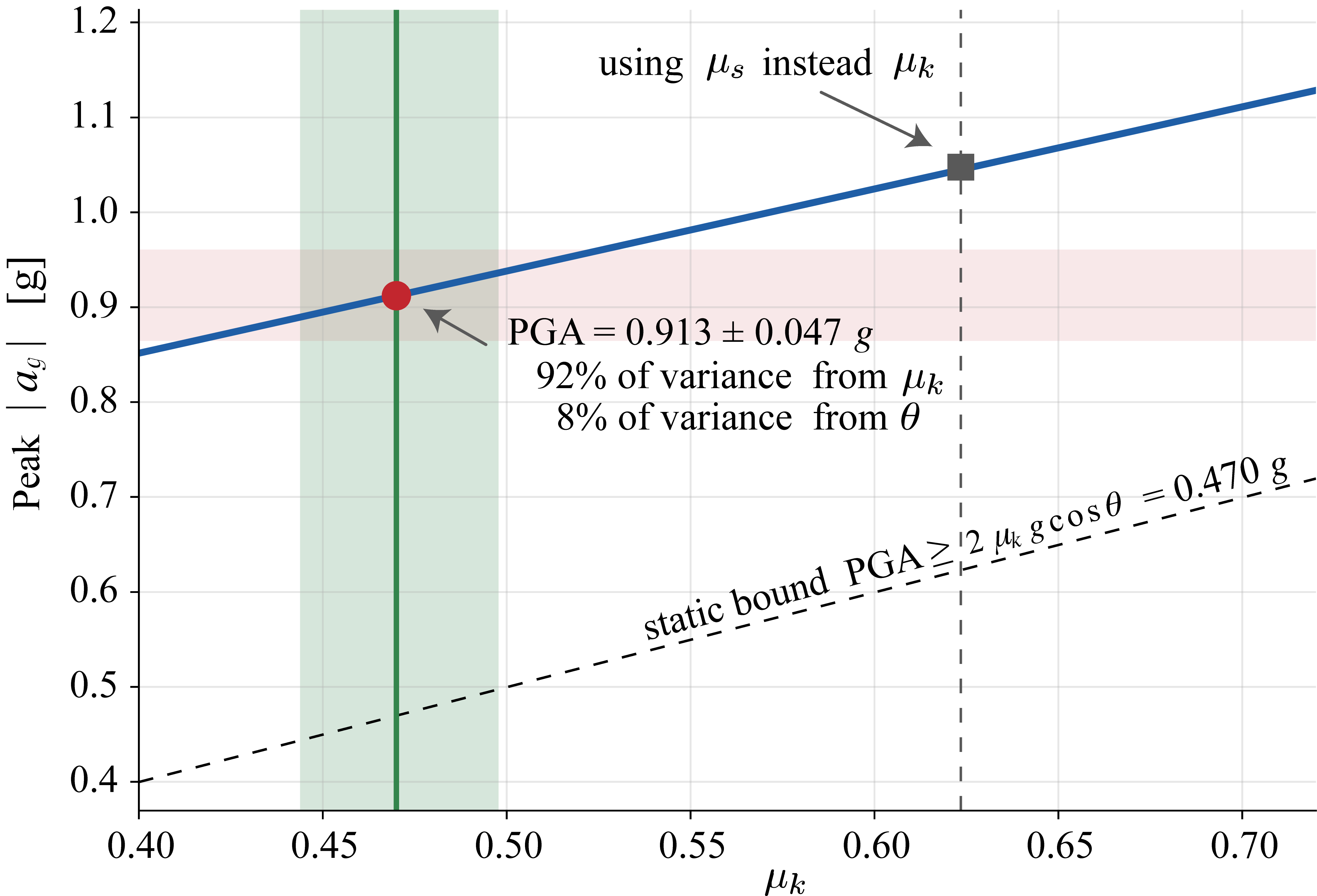}
  \caption{Propagation of the kinetic coefficient to the peak ground acceleration. The friction coefficient accounts for \SI{92}{\percent} of the variance in \cref{eq:pga}, and inclination for the remaining \SI{8}{\percent}. The vertical band is $\mu_{k}$ from \cref{eq:muk} and the horizontal one is the resulting peak. The marker at the breakaway value of \cref{eq:mus} shows the cost of using the static coefficient in the inverse. The dotted line is the model-independent bound \cref{eq:cota}, which the inversion exceeds everywhere because the post was observed to accelerate and not merely to break away.}
  \label{fig:sensibilidad}
\end{figure}

The dispersion is dominated by the interface to the near exclusion of everything else. Freezing $\theta$ and varying only $\mu_{k}$ reproduces \SI{0.045}{\gravity} of the \SI{0.047}{\gravity}; freezing $\mu_{k}$ and varying only $\theta$ gives \SI{0.013}{\gravity}, so that the inclination accounts for \SI{8}{\percent} of the variance; and the correlation between the recovered peak and $\mu_{k}$ is $+\num{0.96}$ against $+\num{0.30}$ for the inclination. This is the expected structure, since $\mu_{k}$ enters the inversion multiplicatively, and it carries a clear practical instruction: improving this estimate is a matter of a better bench measurement, not of better video. A dozen more dynamometer readings would do more for the result than a higher-resolution camera. The contrast with the inclination is the point of having surveyed it. An inclination known only to within two degrees would contribute \SI{0.025}{\gravity}; surveyed to \ang{0.92}, it contributes about half of that. \Cref{fig:sensibilidad} shows why the decomposition comes out this way. The recovered peak is an affine function of the kinetic coefficient, so the coefficient does not merely dominate the variance, it sets the answer through a law with no curvature in it and the propagation could have been done in closed form. The same figure prices the distinction of \cref{sec:friccion} between the two coefficients: substituting the breakaway value \cref{eq:mus} for \cref{eq:muk}, which is the natural mistake when only a dynamometer is available, moves the peak to \SI{1.048}{\gravity}, high by \SI{15}{\percent} and well outside the interval of \cref{eq:pga}. That error is three times the whole propagated uncertainty, and it is a modeling error rather than a measurement one, so no amount of additional
bench work would reveal it.

The model-independent bound \cref{eq:cota} gives
$\mathrm{PGA}\ge\SI{0.470}{\gravity}$ from the bare fact that the post moved. The Monte Carlo minimum over the twenty thousand samples is \SI{0.720}{\gravity} and its fifth percentile is \SI{0.837}{\gravity}, so the whole distribution sits above the bound, which is the consistency one would want between the two routes to a number.

The same twenty thousand samples were carried through the integration of \cref{sec:recuperado}, and the result is the marker drawn at $f_{c}=\SI{0.70}{\hertz}$ in \cref{fig:mc}(f): a peak ground velocity of \SI{898}{\milli\meter\per\second} with a standard deviation of \SI{57}{\milli\meter\per\second}, a coefficient of variation of \SI{6.3}{\percent}. Taken alone that figure would suggest the velocity is as well determined as the acceleration. It is not, and the panel is drawn to make the comparison impossible to avoid: the same curve spans a factor of \num{2.2} across the three cut-off frequencies that \cref{fig:determinada} defends and a factor of \num{5.4} between \SIrange{0.50}{1.50}{\hertz}, without ever flattening.

\subsection{What the Index Would Have Cost Here}
\label{sec:costo}

The identity \cref{eq:identidad} can be evaluated directly on this record, without reference to any synthetic case, because it depends only on the sliding direction and the interface. Over the strong-motion window the post slides on \num{40} of the \num{49} steps, and across those steps the sliding direction turns by \ang{10.1} from one step to the next on average, with a median of \ang{9.6} and a largest value of \ang{22.2}. The discrepancy between the two inverses is accordingly \SI{0.099}{\gravity} in root mean square over the sliding steps and \SI{0.112}{\gravity} over the window as a whole, with a largest single-step value of \SI{0.470}{\gravity} at $t=\SI{10.44}{\second}$ against the bound $2\mu_{k}g\cos\theta=\SI{0.939}{\gravity}$. It is not an artifact of the stick-slip transitions: the window contains exactly one, and removing it together with its immediate neighbors leaves \SI{0.096}{\gravity}, so \SI{97}{\percent} of the discrepancy lives in sustained sliding.

The comparison that settles the importance of this is not with the measurement noise but with the reported uncertainty. Inverting the continuous equation instead of the scheme moves the recovered peak from \SI{0.912}{\gravity} to \SI{0.839}{\gravity} and shortens the re-integrated travel by \SI{17}{\percent}, from the \SI{449}{\milli\meter} of \cref{sec:recuperado} to \SI{372}{\milli\meter}, turning a small overshoot of the measured slip into a substantial shortfall. The displacement of the peak, \SI{0.073}{\gravity}, is larger than the entire dispersion of \cref{eq:pga}, $\pm\SI{0.047}{\gravity}$, which this paper obtains by characterizing the interface on the bench and surveying the slab with a level. The index is therefore not a refinement to be weighed against the measurement uncertainties. It is larger than all of them together.

On this record the usual acceptance check does register the error, but it registers it as a smaller thing than it is. Comparing the relative acceleration that the forward simulation reproduces against the measured one, the exact inverse leaves \SI{0.046}{\gravity} in root mean square and the incorrect one \SI{0.096}{\gravity}, so an analyst watching that residual would see it double and would know that something was wrong. What the residual does not convey is which way the error runs, or how large it is in the quantity actually being reported: the same doubling accompanies a peak low by \SI{8}{\percent} and a travel short by \SI{17}{\percent}, because the discrepancy is a bias and a root-mean-square residual carries magnitude without sign. An integrated quantity such as the recovered travel states the size and the direction of the error rather than merely its presence, which is why the forward check of \cref{sec:recuperado} is reported in displacement and not in acceleration.

\section{Conclusions}
\label{sec:conclusiones}

For a rigid body sliding in two dimensions under Coulomb friction with unknown base excitation, we have shown the following.

The Moreau time stepper admits an exact inverse, \cref{eq:inverso}, which is explicit, local, vectorizable and free of regularization. Given the sampled relative velocity and the interface parameters, it returns the ground acceleration over each sliding step with no approximation, and re-integration reproduces the input trajectory to machine precision.

Inverting the continuous equation of motion instead, which evaluates the friction direction at the beginning of the step rather than at its end, is not a minor discretization choice. The two inverses differ by the exact identity $\Delta\ag=\mu_{k}g\cos\theta(\evk{i+1}-\evk{i})$, bounded by $2\mu_{k}g\cos\theta$, which is of order $1\,g$ for a typical interface, and given in closed form by $2\mu_{k}g\cos\theta\sin(\Omega_{v}\Delta t/2)$ under constant rotation. The implicit direction is not optional: it is the mechanism by which the scheme sticks, and an explicit direction is incompatible with exact sticking.

The effect is invisible in one dimension, where the sliding direction is piecewise constant, and active at essentially every step in two, where it rotates. It is not a transition artifact: the strong-motion window of the field record contains a single stick-slip transition, and removing it leaves \SI{97}{\percent} of the discrepancy in place.

The discrepancy is a bias, not noise. Sweeping tracking noise over two orders of magnitude, the root-mean-square errors of the two inverses cross, and above the crossing the incorrect inverse has the lower root-mean-square error, while its recovered travel remains roughly a tenth short at every noise level. Root-mean-square agreement in acceleration is therefore not an acceptance criterion for this class of inverse problem; an integrated quantity such as recovered displacement is, because it is sensitive to bias and insensitive to zero-mean noise.

A complete chain from an incidental surveillance recording of the 2026 La Guaira earthquake to a ground acceleration with a stated uncertainty is demonstrated and reported stage by stage. Camera egomotion is removed against a datum that belongs to the ground rather than to the camera, and the removal is validated to \SI{1.203}{\pixel} in root mean square; the scene is calibrated from vanishing points recovered from the built environment, with scale fixed by one known object height; the sliding body is tracked under partial occlusion, with the unreliable intervals identified from the template correlation itself and bridged rather than inverted; and the interface is characterized on the bench, and the slab is surveyed with a spirit level so that neither the inclination nor its azimuth is left to assumption. The exact inverse recovers a peak ground acceleration of \SI{0.912}{\gravity} from \SI{428.9}{\milli\meter} of measured slip, and propagation of the interface and inclination uncertainties gives $\mathrm{PGA}=\num{0.912}\pm\SI{0.047}{\gravity}$, a coefficient of variation of \SI{5.1}{\percent}, in a direction of
\ang{133.4}. The peak ground velocity is not recovered, for reasons
given in \cref{sec:recuperado} that rest on the duration of the slip
rather than on the inversion: observing only the delineator's own
motion over a two-second slide bounds that motion from below but does
not resolve the longer periods that set a ground velocity, which is a
limitation of this record and not of the method in general.

For seismic engineering, this supports three statements. A displaced rigid body is a usable instrument and not merely a damage indicator, since the dispersion it delivers is set by the interface characterization rather than by the imagery, and the relevant comparison is not against a free-field accelerograph but against having no measurement at the site at all. The lower bound $\mathrm{PGA}\ge\mu_{k}g\cos\theta=\SI{0.47}{\gravity}$ is the robust product of the analysis, because it uses no imagery, no calibration, no tracking and no time step, and it is worth collecting systematically in post-earthquake reconnaissance wherever an object of known interface is found displaced. Finally, two dimensions change the character of sliding-block analysis rather than refining it: any two-dimensional extension of Newmark analysis, forward or inverse, inherits the index bias quantified here, and correcting it costs one evaluation of the sliding direction at the end of the step rather than at its beginning.

\section*{Acknowledgements}
The authors gratefully acknowledge Dr. Ronald Torres, MSc. Ahmad Idrees, MSc. José Rengel, and the entire team of the Instituto de Materiales y Modelos Estructurales (IMME), Universidad Central de Venezuela, for their valuable logistical support and assistance during the post-earthquake field reconnaissance campaign. Friction and slope data were initially estimated from the camera model and subsequently verified during a field mission by the Department of Civil Engineering at the University of Chile. Additionally, Dr. Ruiz’s startup package at the University of Michigan-Dearborn partially funded the post-processing of the data.


\bibliographystyle{model1-num-names}
\bibliography{referencias_mssp}

\end{document}